\documentclass[]{aastex}
\usepackage{lineno}

\def\wfirst{\textit {Roman}}  

\def\Lstar{L_{\star}}
\def\REarth{r_\oplus}

\def\LSun{L_\odot}

\def\gapp{\lower 3pt\hbox{${\buildrel > \over \sim}$}\ }
\def\lapp{\lower 3pt\hbox{${\buildrel < \over \sim}$}\ }
\def\proptosim{\lower 3pt\hbox{${\buildrel \propto \over \sim}$}\ }

\def\arcsec{$^{\prime\prime}$}
\def\degree{$^\circ$}

\begin{document}

\title{Starshade Rendezvous: \\
  Exoplanet Orbit Constraints from Multi-Epoch Direct Imaging}


\author{Andrew~Romero-Wolf}
\affiliation{Jet Propulsion Laboratory, California Institute of Technology, Pasadena, CA 91109, USA}

\author{Geoffrey~Bryden}
\affiliation{Jet Propulsion Laboratory, California Institute of Technology, Pasadena, CA 91109, USA}

\author{Greg Agnes}
\affiliation{Jet Propulsion Laboratory, California Institute of Technology, Pasadena, CA 91109, USA}

\author{Jonathan W. Arenberg}
\affiliation{Northrop Grumman Aerospace Systems, Redondo Beach, CA 90278, USA}

\author{Samuel Case Bradford}
\affiliation{Jet Propulsion Laboratory, California Institute of Technology, Pasadena, CA 91109, USA}

\author{Simone D'Amico}
\affiliation{Stanford University, Stanford, CA 94305, USA}

\author{John Debes}
\affiliation{Space Telescope Science Institute, Baltimore, MD 21218, USA}

\author{Matt Greenhouse}
\affiliation{NASA Goddard Space Flight Center, Greenbelt, MD 20771, USA}

\author{Renyu Hu}
\affiliation{Jet Propulsion Laboratory, California Institute of Technology, Pasadena, CA 91109, USA}

\author{Steve Matousek}
\affiliation{Jet Propulsion Laboratory, California Institute of Technology, Pasadena, CA 91109, USA}

\author{Jason Rhodes}
\affiliation{Jet Propulsion Laboratory, California Institute of Technology, Pasadena, CA 91109, USA}

\author{John Ziemer}
\affiliation{Jet Propulsion Laboratory, California Institute of Technology, Pasadena, CA 91109, USA}



\begin{abstract}
The addition of an external starshade to the {\it Nancy Grace Roman Space Telescope} will enable the direct imaging of Earth-radius planets orbiting at $\sim$1 AU. Classification of any detected planets as Earth-like requires both spectroscopy to characterize their atmospheres and multi-epoch imaging to trace their orbits. We consider here the ability of the Starshade Rendezvous Probe to constrain the orbits of directly imaged Earth-like planets. The target list for this proposed mission consists of the  16 nearby stars best suited for direct imaging. The field of regard for a starshade mission is constrained by solar exclusion angles, resulting in four observing windows during a two-year mission. We find that for habitable-zone planetary orbits that are detected at least three times  during the four viewing opportunities, their semi-major axes are measured with a median precision of 7 mas, or a median fractional precision of 3\%. Habitable-zone planets can be correctly identified as such 96.7\% of the time, with a false positive rate of 2.8\%. If a more conservative criteria is used for habitable-zone classification (95\% probability), the false positive rate drops close to zero, but with only 81\% of the truly Earth-like planets correctly classified as residing in the habitable zone.
\end{abstract}
\keywords{planets, imaging}


\section{Introduction}

The Starshade Rendezvous Probe (SRP) mission concept proposes adding a Starshade to the {\it Nancy Grace Roman Space Telescope} enabling the detection of habitable zone exoplanets and characterization of their atmospheres~\citep{seager19}. \cite{paper1} described in detail the technical basis for the SRP study report~\citep{seager19} along with the simulations used to estimate sensitivity of the observatory. The corresponding software is publicly available for reproduction of the results below and for comparison with similar simulations.\footnote{\url{https://github.com/afromero/Starshade_Rendezvous_Probe_sims}}. The main result of these studies is that SRP is capable of discovering Earth-size planets in the habitable zones of nearby stars using the relatively moderate aperture of the \wfirst~space telescope~\citep{paper1} along with the Coronagraph Instrument (CGI). 

While the SRP science objectives include quantifying the amount of habitable zone dust around nearby stars and measuring the metallicity of known gas giant planets, the primary driver is the detection and characterization of Earth-like planets. The overall strategy, described in more detail in \citet{paper1}, involves three main steps:
1) initial detection via direct imaging,
2) habitable zone determination via orbit tracing, and
3) atmosphere characterization via spectroscopy.
The integration times necessary to image and to take spectra of Earth-like planets (steps 1 and 3) were taken into account with a model of the observatory. However, step 2 is more complicated since the observatory field of regard is constrained by solar exclusion angles, typically limiting the target observing windows to two $\sim$30 day blocks per year -- a total of 4 observing opportunities during the assumed 2-year mission lifetime. Depending on the orientation and phase of a planet's orbit, it may or may not be visible during each of these 4 observing windows. In \citet{paper1}, we assumed that detecting the planet during at least 3 of the 4 epochs would be sufficient to determine if a planet lies within its parent star's habitable zone. In this paper, we consider this step in more detail, performing multi-epoch orbit fitting for the target list and expected signal-to-noise given by the \citet{paper1} observatory model.

Measurement of planetary orbits has been previously modeled for several types of observations -- radial velocity, astrometric wobble, and coronagraphic direct imaging. Examples include: 
1) \citet{mawet19} combining radial velocity measurements with direct imaging upper limits to improve the orbital fit for the planet eps Eri b,
2) \citet{ford06} quantifying the robustness of planetary orbit determination via the stellar astrometric signal, concentrating on the difficulty posed by multi-planet systems, and
3) \citet{guyon13} considering simultaneous stellar astrometry and direct imaging, finding that Earth-like planets can be characterized in just a few observations. 
Among the studies that, like this paper, concentrate on direct imaging, \citet{blunt17} performed a detailed analysis of orbital constraints based on the small fraction of an orbit that is traced by known long-period direct-imaged planets. For theoretical cases where the observations span at least half an orbital period, \cite{horning19} find that three equally spaced observations with SNR $\geq$ 10  can measure the semi-major axis and eccentricity to 10\%. \cite{guimond19} also consider direct imaging of shorter period planets, finding that a habitable zone planet's semi-major axis can be measured to within 5\% if it is observed with precision 3.5 mas over three epochs each spaced by at least 90 days apart.

Here we consider the results obtainable by a specific mission concept -- the Starshade Rendezvous Probe.
Unlike previous work, this includes:
1) a realistic signal-to-noise calculation as a function of stellar illumination, rather than an assumed astrometric precision, 
2) a starshade-specific inner working angle that obscures planet images close to the star, 
and
3) observing windows based on known target sky positions and observatory pointing constraints, 
not just arbitrarily spaced images. 
Furthermore we focus here not on general orbit fitting results, but on a specific science question -- 
whether or not we can determine if a planet lies in its parent star's habitable zone.

In the following, we first summarize the observatory model from~\cite{paper1} used to calculate the 
signal-to-noise for each planet image \S\ref{model}. For many sets of simulated observations, we then extract orbital elements for each injected planets \S\ref{orbitfit}. We give the resulting precision for the orbital fits in \S\ref{results} and 
summarize in \S\ref{summary}.


\section{Observing Model}\label{model}

We briefly summarize the models used for planet properties, orbit propagation, and the observatory. A detailed presentation of these models can be found in \citet{paper1}.

\subsection{Targets}
Planet sizes and orbital periods are drawn randomly from these defined ranges for Earth-like planets, based on the distribution defined by SAG-13~\citep{belikov17} and modified by HabEx to include the dependence of the orbital semi-major axis on the lower limit of planet radii. The orbital period $P$ defines the orbital radius $a_{p}$ by way of the stellar mass $M_\star$ using Kepler's third law. For sampling of Earth-like exoplanets, the orbits are assumed to be circular, consistent with most previous studies, e.g.~\citet{stark16}. However, when fitting, we allow eccentricity to be a free parameter (see \S3). 

The orbital radii are sampled over a range from inside the inner habitable zone (defined as 0.95$\sqrt{L_\star}$, where $L_\star$ is the stellar luminosity relative to the Sun) to outside the outer habitable zone \citep[defined as 1.67$\sqrt{L_\star}$;][]{kasting93}.

The range of planet radii considered is bounded above at $r_{pl}\leq 1.4$ $\REarth$, based on evidence that suggests that planets with radius below  are predominantly rocky \citep{rogers15}. The lower bound on terrestrial planet radii depends on the planet's ability to retain an appreciable atmosphere, which, in turn depends on their stellar illumination. This results in a dependence on the planet's semi-minor axis $a_{\rm p}$, modified by the stellar luminosity to give $r_{pl} \geq 0.8 a^{1/2}_{p} / L^{1/4}_{\star}$ \citep{zahnle17}. 

We model Earth-like exoplanets assuming the scatter light isotropically using a Lambertian illumination phase function with a geometric albedo of 0.2 \citep{robinson18}. We model the star as a blackbody radiator with the parameters provided in ExoCat~\citep{turnbull12}. We include obscuring dust, both in the target system (exozodiacal dust with a fiducial value of 4.5~zodi~\citep{ertel20}) and locally in the Solar System (zodiacal dust using the model of~\citet{leinert98}).
Assumed planet and dust characteristics are summarized in Table \ref{planetParamTable}.

\begin{deluxetable}{l|c}
\tablewidth{8.5in}
\tablecaption{Planet and Dust Parameters
\label{planetParamTable}}
\tablehead{Parameter & Value (or [Range])}
\startdata
Earth-like planet geometric albedo$^a$ & 0.2 \\
Earth-like planet radius$^b$  &  $[0.8 a_{\rm p}^{0.5}-1.4]$ $\REarth$ \\
Habitable zone  & $[0.95 - 1.67] \sqrt{L_{\star}}$ \\
\hline
Zodiacal dust brightness & \citet{leinert98}  \\
Exozodi dust brightness$^c$ &  4.5 zodi \\
\hline
\enddata
\tablenotetext{a}{For the assumed isotropic scattering, this geometric albedo
  is equivalent to 0.3 spherical albedo.}
\tablenotetext{b}{The semi-major axis $a_{\rm p}$ is modified by a factor of $\sqrt{L_\star}$ to account for stellar irradiance.}
\tablenotetext{c}{The unit of 1 zodi is equivalent to 22 mag/arcsec$^2$.}
\end{deluxetable}

While the use of an occulting starshade does allow for detection of fainter and closer planets, it also puts constraints on the
allowed times of observation. The telescope-starshade system has a region of allowed Sun angles over which it can operate with the lower limit defined by the exclusion angle from the baffle of the telescope and the outer limit defined by reflection and scattering of sunlight off the starshade into the telescope baffle (Table \ref{missionParamTable}).

\subsection{Observatory}

The integration times are based on the Starshade/\wfirst\ system parameters provided in Table~\ref{missionParamTable}. \wfirst\ has a telescope diameter of $2.4$~m resulting in a point spread function of 65~mas at 750 nm wavelength. We assume observations in the 615 -- 800~nm band with an end-to-end efficiency, including optical throughput and detector efficiency, of 3.5\% in imaging mode. The Starshade has an inner working angle (IWA) of 100~mas, below which planets are assumed to not be observable. 
The instrument contrast at the IWA and above is assumed to be $4\times10^{-11}$, as calculated by \citet{seager19}.
Integration times vary between 1 -- 6.3 days, depending on the target (explained below).

\begin{figure}[ht]
\begin{center}
    \includegraphics[width=7in]{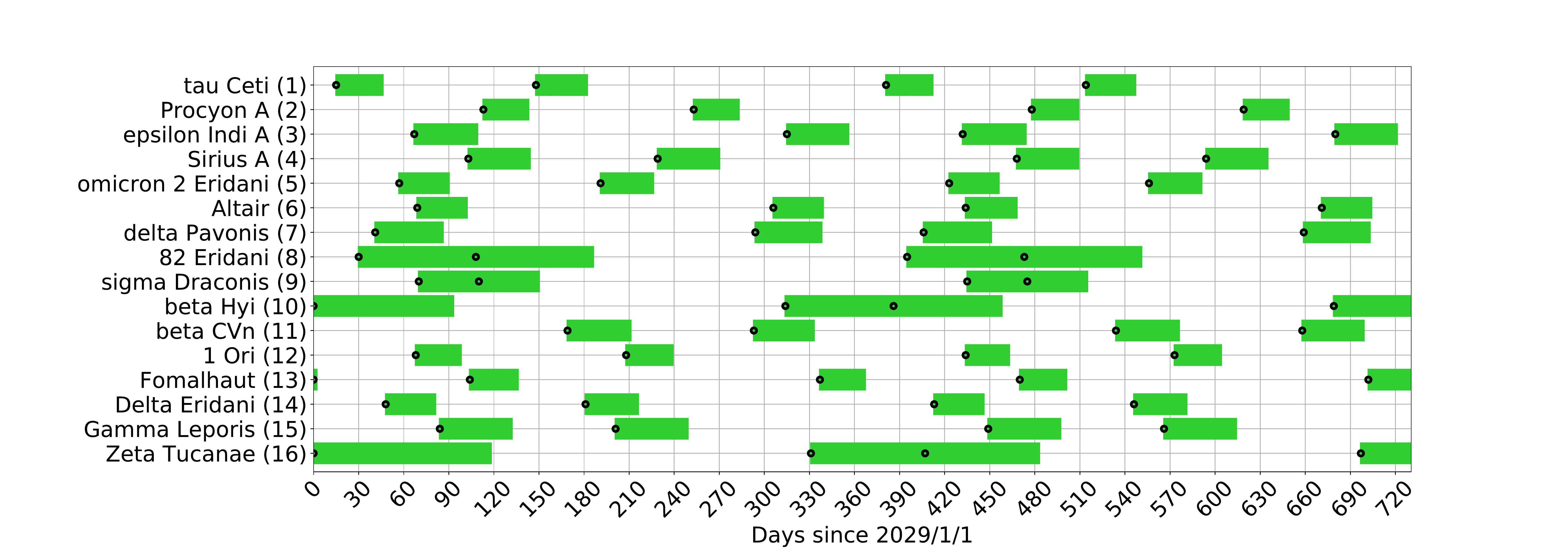}
  \end{center}\caption{
    Target star observing windows (reproduced from~\citet{paper1}) resulting from the telescope and starshade solar exclusion angles. Each target typically has two $\sim$30-day-long observing windows per year.  Targets at high ecliptic latitude can have longer observing windows per year.  The black dots mark the four desired observation start times in a two-year period. This is driven by the need to allow for sufficient time to spectrally characterize a planet if it is bright enough.  
  }\label{habitability_windows}
\end{figure}

Besides the sensitivity, determined by the parameters given above, the main constraints to orbit reconstruction are the nominal mission lifetime of 2~years and the solar exclusion angles. The minimum solar exclusion angle of $54^\circ$ is determined by the telescope baffle while the maximum solar exclusion angle of $83^\circ$ is determined by scattering off the edge of the Starshade. 
For \wfirst's orbit around the L2 Sun-Earth Lagrange point, the calculated observing windows are shown in Figure~\ref{habitability_windows}. During the 2-year lifetime of the mission, there are generally 4 opportunities to observe each target. It is assumed that targets with Earth-like planet candidates, following the decision tree laid out in \cite{paper1}, will be visited once in each of the 4 observing windows available. While spectral characterization can be performed with only a single visit with favorable illumination phase,  multiple epochs are needed to constrain the planet's orbit, in particular its semi-major axis, which determines whether the planet is in the habitable zone. To best trace the orbit, the timing of the observations should be as evenly spaced as possible over the orbital period, given the limited observing windows and mission lifetime.

\begin{deluxetable}{l|c}
\tablecaption{Mission Parameters
\label{missionParamTable}}
\tablehead{Parameter & Assumed Performance}
\startdata
Mission lifetime         & 2 years \\
Telescope primary mirror & 2.4 m \\
Imaging resolution & 65 mas at 750 nm \\
Imaging bandpass & 615 -- 800 nm \\
Imaging end-to-end efficiency & 0.035 \\
\hline
Solar exclusion angle (min)  & 54\degree \\
Solar exclusion angle (max)  & 83\degree \\
\hline
Inner working angle (IWA) & 100 mas \\
Instrument contrast & $4 \times 10^{-11}$ \\
\hline
Imaging integration time & 1 -- 6 days \\
\enddata
\end{deluxetable}

The properties of the target stars are shown in Table~\ref{targetList}. In \citet{paper1} we defined the single-visit completeness as the probability that an Earth-like planet would be detected during one target observation at a random time. The imaging integration time for each target is set by identifying what is required to reach a single-visit completeness of at least 50\% with a minimum value of 1 day and a maximum value of 6.3 days
(see Table~\ref{targetList}).
We also defined the {\it orbital completeness} as the probability that a randomly selected orbit will be detectable (S/N $\geq$ 7) during at least 3 of its 4 observing epochs.
For most of the targets about half of the planet orbits meets this criterion,
but the orbital completeness can sometimes fall below 20\% for
systems where the planet signal is relatively weak
(again, see Table~\ref{targetList}).
In the next section, we calculate whether 3 detections in 4 visits is sufficient
to determine whether a planet lies within its habitable zone.

\setlength{\tabcolsep}{4pt} 
\rotate
\begin{deluxetable}{l|cccccc|c||cc|cc|cc|cc}
\tabletypesize{\scriptsize}
\tablecaption{Astrometric precision for Earth-like planets 
\label{targetList}}
\tablehead{
\multicolumn{5}{c}{} & 
\multicolumn{2}{c}{Habitable Zone} &
\multicolumn{3}{c}{} & 
  \multicolumn{2}{c}{Astrom.\ Precision$^e$} &
  \multicolumn{4}{c}{Habitable Zone classification} \\
Star Name & Distance & $V$ & $\Lstar$ & $M_{\star}$ & $a_p$  & Period &
  Int.\ Time & \multicolumn{2}{c}{Completeness} &
  per epoch &  $a_{p}$ & 
 \multicolumn{2}{c}{50\% Threshold} & \multicolumn{2}{c}{95\% Threshold} \\
 & (pc) & (mag) & ($\LSun$) & ($M_\odot$) & (mas) & (years) &
  (days) & Single-visit & Orbital$^d$  & (mas) & (mas) & 
    False Pos. & False Neg. & False Pos. & False Neg.}
\startdata
tau Ceti$^{bc}$& 3.7 & 3.5 & 0.52 & 0.80 & 187 -- 329 & 0.64 -- 1.5 & 1.0 & 0.67 & 0.55 & 3.2 & 5.7  & 2.2\% & 2.0\% & 0.0\% & 12.6\% \\
Procyon$^a$    & 3.5 & 0.4 & 7.1  & 1.49 & 722 -- 1270& 3.3 -- 7.7 & 1.0 & 0.65 & 0.54 & 3.2 & 28.2 & 2.2\% & 1.5\% & 0.0\% & 14.1\% \\
eps Ind$^{ac}$ & 3.6 & 4.7 & 0.23 & 0.68 & 124 -- 219 & 0.36 -- 0.84 & 2.5 & 0.67 & 0.54 & 2.3 & 3.5  & 1.7\% & 0.5\% & 0.0\% &  7.7\% \\
Sirius$^a$    & 2.6 &$-$1.4& 30.5 & 2.40 &1993 -- 3503& 7.6 -- 17.8 & 1.0 & 0.58 & 0.53 & 3.9 & 72.5 & 3.5\% & 2.1\% & 0.0\% & 16.9\% \\
\hline
omi 2 Eri$^c$  & 5.0 & 4.4 & 0.42 & 0.81 & 124 -- 218 & 0.54 -- 1.3 & 6.3 & 0.65 & 0.52 & 2.5 & 4.7  & 0.8\% & 1.5\% & 0.2\% & 9.1\% \\
Altair         & 5.1 & 0.8 & 10.7 & 1.83 & 605 -- 1064& 4.0 -- 9.4 & 2.5 & 0.58 & 0.51 & 3.9 & 36.3 & 3.5\% & 3.9\% & 0.0\% & 24.3\% \\
del Pav        & 6.1 & 3.5 & 1.3  & 0.99 & 179 -- 315 & 1.2 -- 2.7 & 6.3 & 0.64 & 0.55 & 3.5 & 4.9  & 2.2\% & 2.2\% & 0.0\% & 12.7\% \\
82 Eri$^c$     & 6.0 & 4.3 & 0.69 & 0.85 & 130 -- 229 & 0.75 -- 1.7 & 6.3 & 0.60 & 0.40 & 3.1 & 4.4  & 1.2\% & 2.4\% & 0.2\% & 11.1\% \\
\hline
sig Dra        & 5.8 & 4.7 & 0.44 & 0.80 & 109 -- 181 & 0.56 -- 1.3 & 6.3 & 0.55 & 0.42 & 3.1 & 4.5  & 1.7\% & 1.1\% & 0.0\% &  7.2\% \\
bet Hyi        & 7.5 & 2.8 & 3.7  & 1.14 & 245 -- 430 & 2.3 -- 5.4 & 6.3 & 0.58 & 0.40 & 3.8 & 20.9 & 4.6\% & 5.2\% & 0.2\% & 31.7\% \\
bet CVn$^a$    & 8.4 & 4.2 & 1.3  & 1.03 & 126 -- 222 & 1.1 -- 2.5 & 6.3 & 0.43 & 0.14 & 4.4 & 4.4  & 2.2\% & 2.3\% & 0.0\% & 10.5\% \\
1 Ori          & 8.1 & 3.2 & 3.0  & 1.24 & 203 -- 358 & 1.9 -- 4.4 & 6.3 & 0.50 & 0.31 & 4.9 & 9.3  & 4.1\% & 5.5\% & 0.0\% & 21.6\% \\
\hline
Fomalhaut$^{ab}$&7.7 & 1.2 & 16.5 & 2.05 & 500 -- 879 & 5.3 -- 12.3 & 6.3 & 0.46 & 0.44 & 4.9 & 50.5 & 4.9\% & 4.0\% & 0.0\% & 40.7\% \\
del Eri        & 9.0 & 3.5 & 3.4  & 1.19 & 193 -- 339 & 2.1 -- 4.9 & 6.3 & 0.46 & 0.26 & 5.3 & 11.6 & 5.7\% & 6.0\% & 0.2\% & 30.2\% \\
gam Lep        & 8.9 & 3.6 & 2.5  & 1.27 & 168 -- 296 & 1.6 -- 3.8 & 6.3 & 0.44 & 0.21 & 5.9 & 8.9  & 1.9\% & 7.2\% & 0.0\% & 30.1\%\\
zet Tuc        & 8.6 & 4.2 & 1.3  & 1.01 & 127 -- 224 & 1.1 -- 2.7 & 6.3 & 0.42 & 0.16 & 5.4 & 4.9  & 2.1\% & 5.4\% & 0.0\% & 24.9\% \\
\hline 
\enddata
\tablenotetext{a}{Binary}
\tablenotetext{b}{Known debris disk}
\tablenotetext{c}{Known to have planet(s)}
\tablenotetext{d}{Orbital completeness requires 3 detections with S/N $\geq$ 7}
\tablenotetext{e}{Astrometric precisions are medians of many sampled planetary orbits}
\tabletypesize{\normalsize}
\end{deluxetable}


\section{Orbit Reconstruction}\label{orbitfit}

Having identified the best targets for detection of Earth-like planets with their observation availability windows, we now  describe our approach to orbit reconstruction. We assume the planet is observed at the beginning of each window as shown in Figure~\ref{habitability_windows}, which provides 4 observing epochs per target in most cases. In cases where the star has a single long availability window per year, we have set the observing times to the beginning and middle of that window. 

For each Monte Carlo sampled planet, we propagate its circular orbit to each of the observing epochs and calculate its illumination phase (see \S2). We apply the observatory model to estimate the planet signal to noise ratio (SNR). Observations with SNR $\geq 7$ are considered to be detections. Otherwise, the observation is rejected as a non-detection. The one-dimensional astrometric uncertainty is approximated according to $\delta \theta = (65~\mathrm{mas})/SNR$. The median astrometric precision for each star ranges from 2.3 to 5.9 mas (see Table~\ref{targetList}). The simulated data are created by taking the true position of the planet and adding two-dimensional Gaussian scatter based on the astrometric precision, $\delta \theta$. 

For the orbit reconstruction, we implemented a forward modeling of Kepler's laws, as described in \cite{mede17}, into our own software package.  We sample all six Keplerian parameters, also including uncertainties in the star's mass and distance, treating them as nuisance parameters.  Table~\ref{fitParamTable} lists the parameters that we fit for each orbit, along with their assumed ranges and prior constraints. We use the \texttt{emcee} Markov chain Monte Carlo (MCMC) software package \citep{foreman-mackey13} to fit the orbit. The MCMC fitting procedure calculates the quality of fit for a series of parameter values, not only converging toward the best set of values but also finding the full parameter ranges that are consistent with the data. Periodic orbital elements ($\omega$, $\Omega$, $T_0$, and $i$) are modulated to stay within their prescribed bounds (typical from 0 to 2$\pi$).

\begin{deluxetable}{c|c|c|c}
\tablecaption{Orbit Fitting Parameters
\label{fitParamTable}}
\tablehead{Parameter & Description & Bounds & Prior Constraint}
\startdata
$a$ & semi-major axis & 0 -- 10 AU & linear \\ 
$e$ & eccentricity &  0 -- 1 & uniform \\
$\omega$ & argument of periastron & 0 -- 360\degree & uniform \\
$i$ & inclination$^a$ & 0 -- 90\degree & $\propto$ sin$(i)$ \\
$\Omega$ & longitude of the ascending node & 0 -- 360\degree & uniform \\
$T_0$ & periastron phase & 0 -- 360\degree & uniform \\
$M_{\star}$ & stellar mass & 0 -- 5 $M_{\odot}$ & observed value with 10\% uncertainty \\
$d_{\star}$ & distance &  0 -- 20 pc & observed value with 1\% uncertainty \\
\hline
\enddata
\tablenotetext{a}{$i=90$\degree\ corresponds to edge-on.}
\end{deluxetable}

\begin{figure}[ht]\begin{center}
    \includegraphics[width=2.3in]{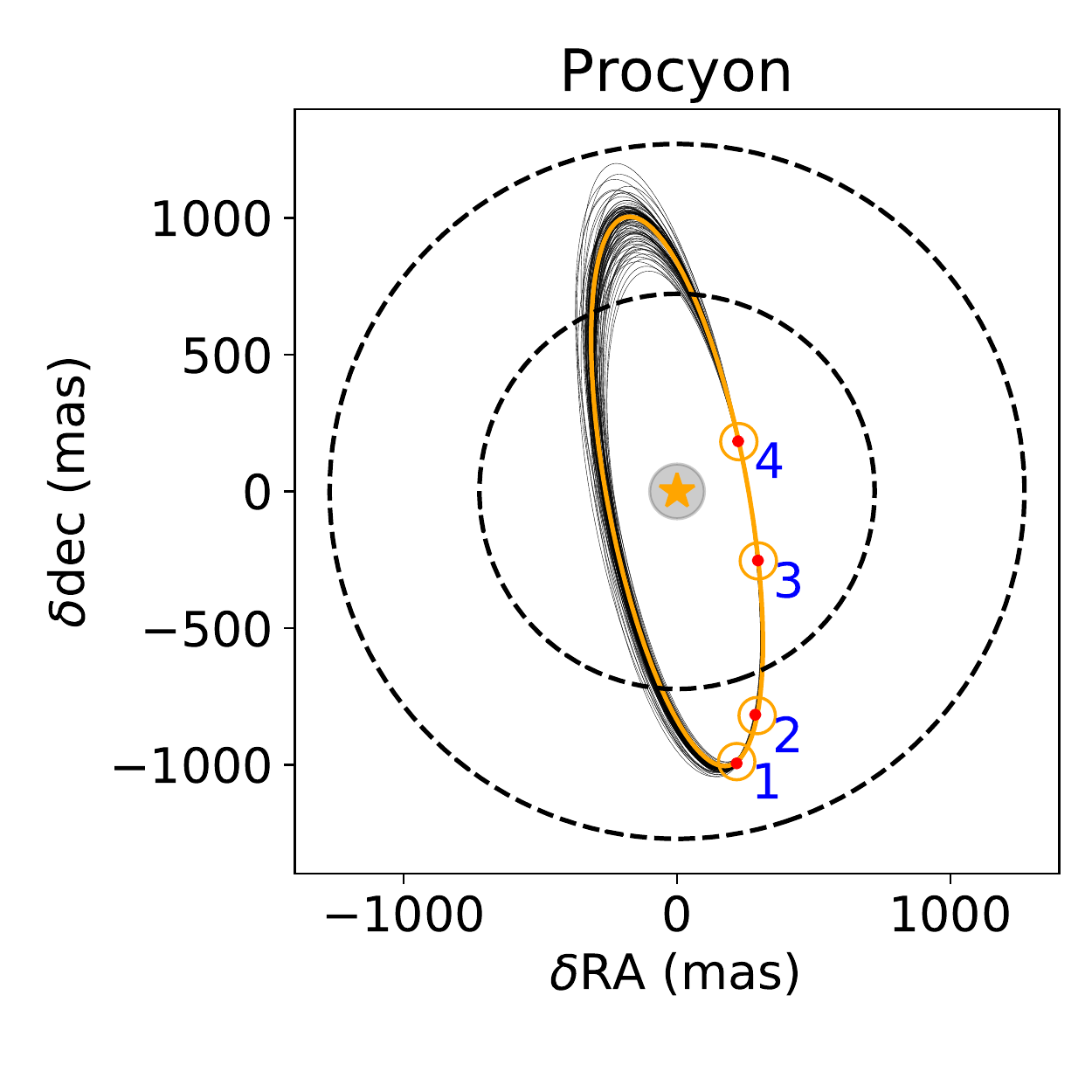}
    \includegraphics[width=2.3in]{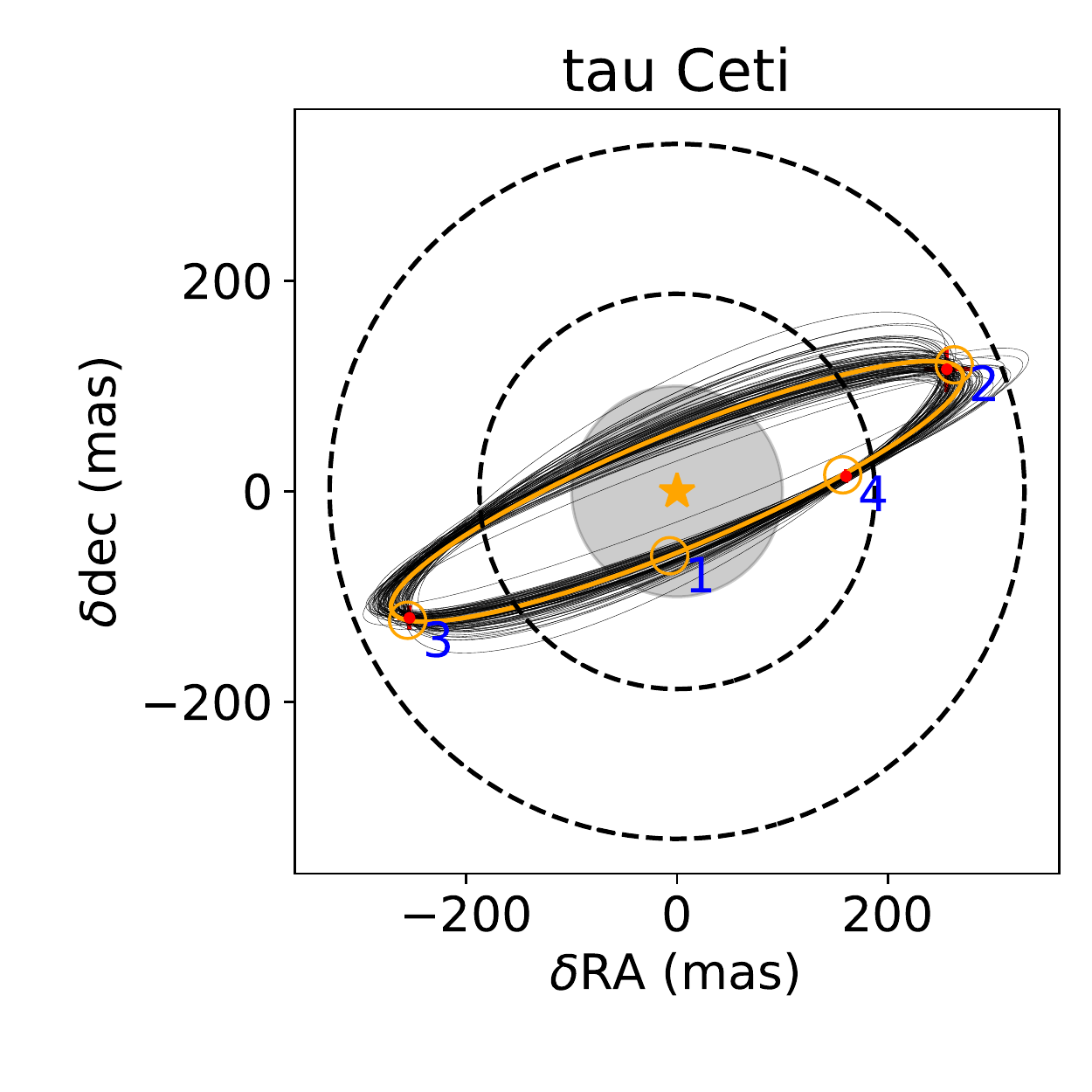}
    \includegraphics[width=2.3in]{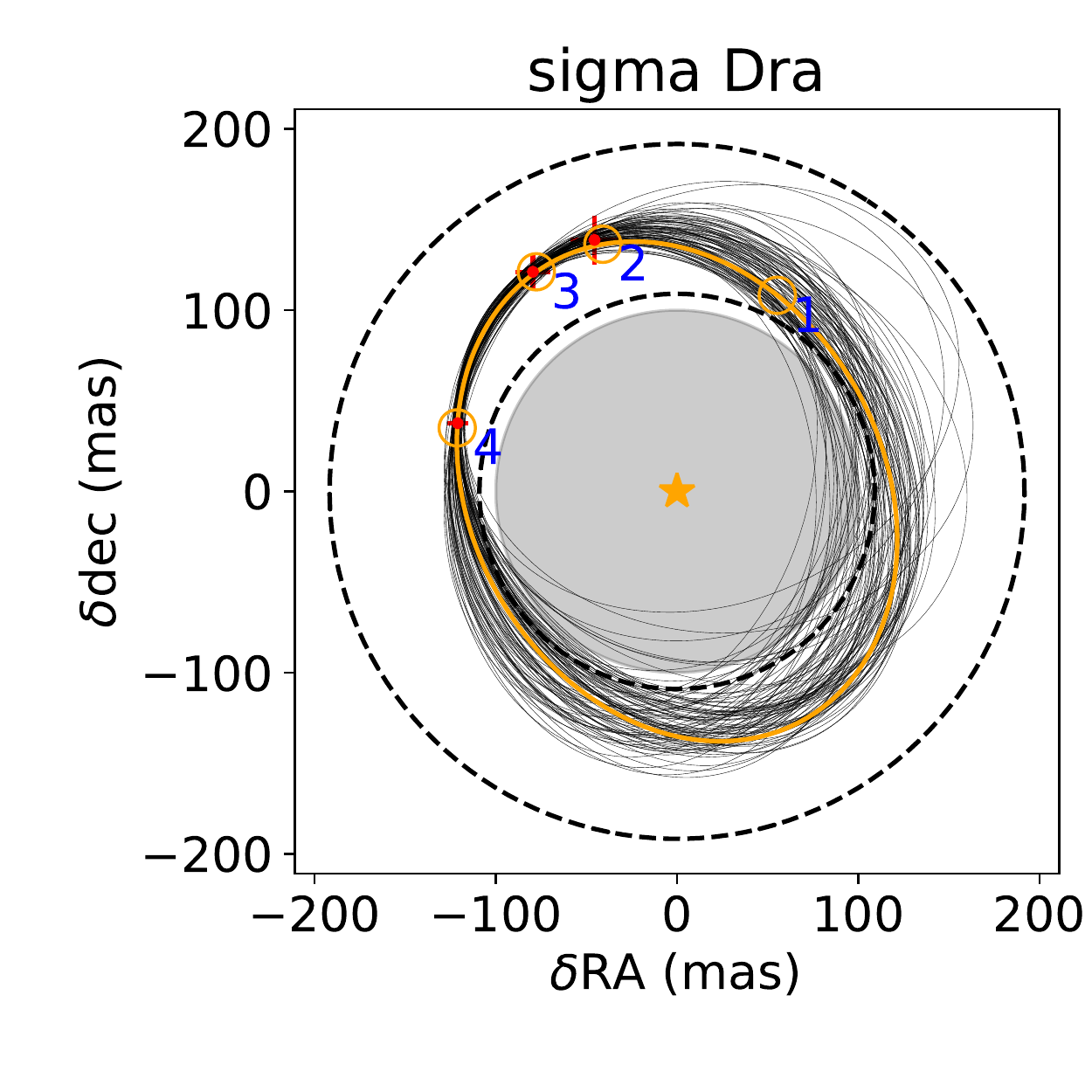}
  \end{center}\caption{
  Example fits are shown for three different stars -- Procyon, tau Ceti, and sigma Dra. In each case, the planet's semi-major axis lies within the habitable zone (dashed lines). The true orbit is shown in orange, with true positions marked as orange circles and simulated observations shown with error bars shown in red only if detected. The numbers indicate the visit number for each of the four observations. Sample best-fit orbits are shown as thin black lines. The region masked by the starshade is shown as a grey circle.
   }\label{orbitFit}
\end{figure}

Three examples of the orbit reconstruction simulations are shown in Figure~\ref{orbitFit}. The first panel shows Procyon, a relatively luminous star (7.1 $\LSun$), meaning that its habitable zone is relatively distant both in angular scale (0.7--1.3\arcsec) and in physical space (2.5--4.5 AU). With a mass of $\sim$1.5~$M_\sun$, planets in the habitable zone have relatively long periods. The randomly selected planet orbiting Procyon is detected in all four observations, but because of the long period, only a fraction of the orbit is traced. The second panel shows a planet orbiting tau Ceti that is only detected in three of the four observing epochs; during the first observation, the planet falls behind the starshade mask (the grey circle in the center of each panel), a common occurrence for planets on inclined orbits. In the third panel (sigma Dra), there are again only 3 successful observations, but in this case the planet is too faint to be detected during the first epoch due to unfavorable illumination phase. Also, because sigma Dra is near the ecliptic north pole, it has only one observing window per year (see Figure \ref{habitability_windows}) and its orbital phase coverage is limited (epoch pairs 1/2 and 3/4 are within the same window). With only three closely-spaced epochs, the fit is relatively poorly constrained.

\begin{figure}[ht]\begin{center}
    \includegraphics[width=4.in]{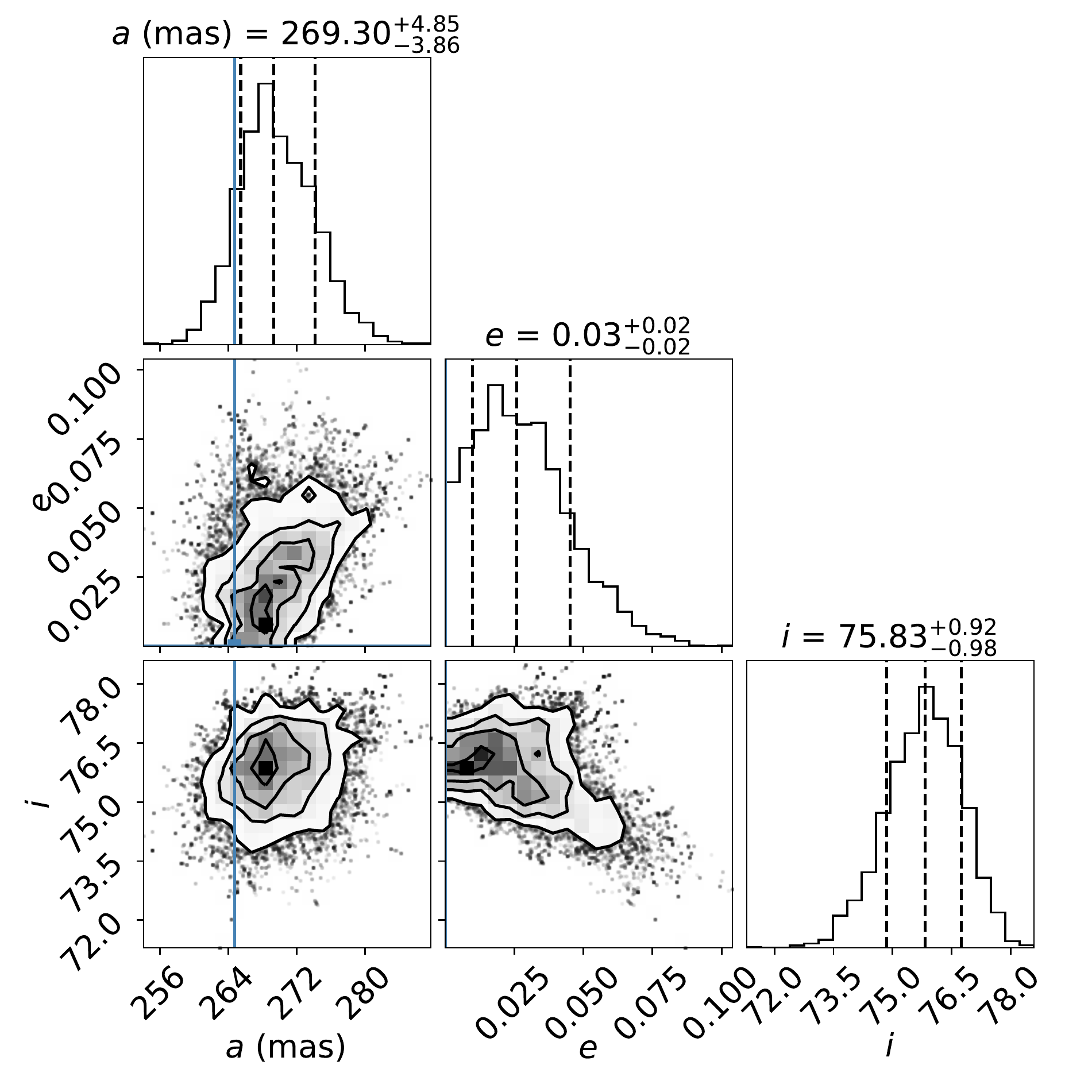}
    \includegraphics[width=2.5in]{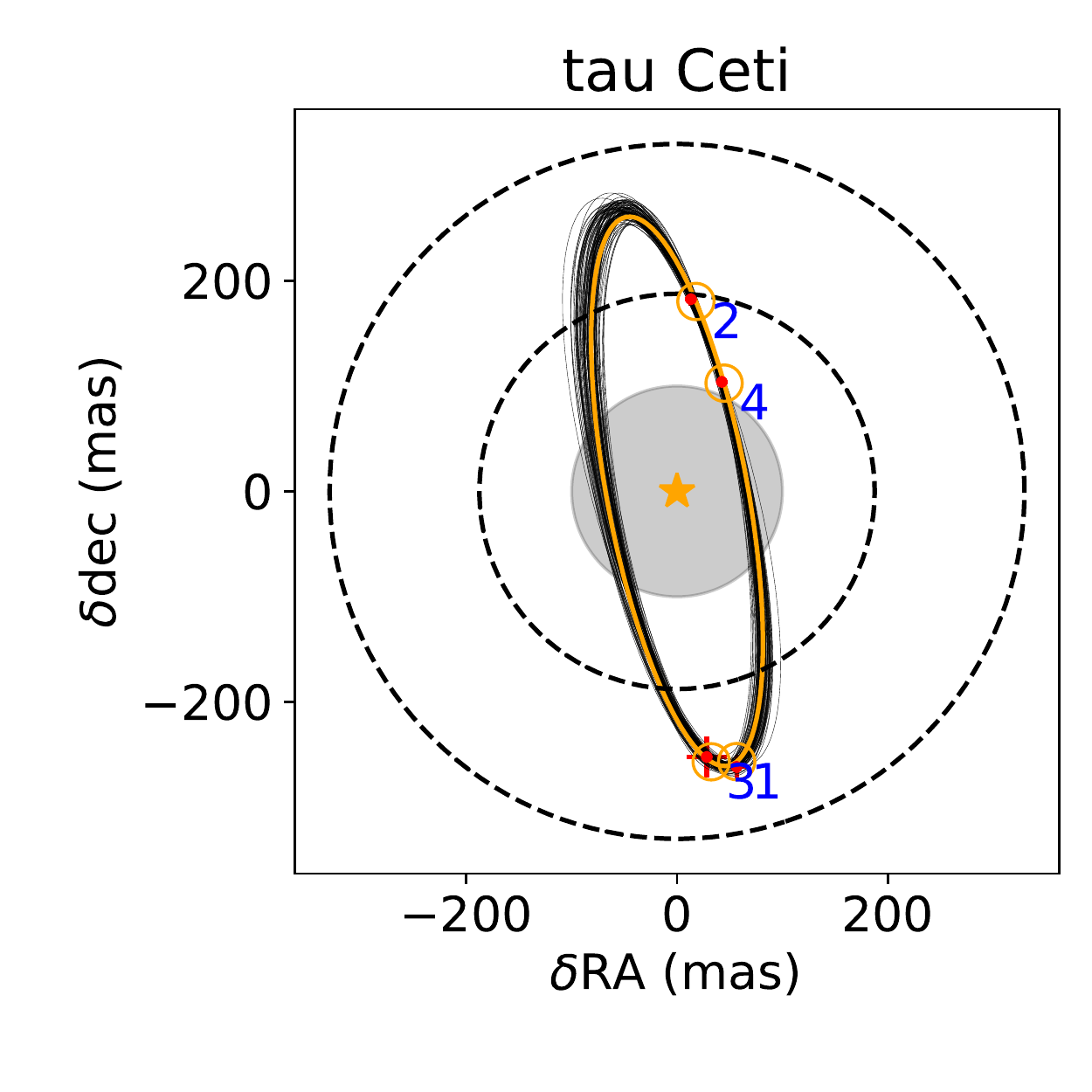}
  \end{center}\caption{
    Best-fit parameters for a random planet orbiting tau Ceti, showing the marginalized probability distributions  for 3 of the 8 model parameters (top histograms) and the correlations between pairs of parameters (central panels, with $1-$, $2-$, and $3-\sigma$ contours). The retrieved parameters are consistent with the true values (blue lines), although the fit eccentricity is necessarily larger than the true circular orbit's. For this example, the planet lies unambiguously inside the habitable zone; the distribution of semi-major axes falls entirely within tau Ceti's 187--329 mas range. 
      }\label{cornerplot}
\end{figure}

An example of the MCMC posterior distributions is shown in Figure~\ref{cornerplot}. For this inclined orbit, the inclination ($i$) and longitude of ascending node ($\Omega$) are well determined and are accurately retrieved.  The retrieved eccentricity ($e$) is necessarily larger than the assumed circular orbit, but is still close to zero (0.03$\pm$0.02). Given the circular orbit, the true argument of periastron ($\omega$) is undefined and the retrieved value is only loosely constrained. Most importantly, the semi-major axis ($a$) is well constrained by the observations, enabling us to determine that the planet lies well within the habitable zone.

\section{Results}\label{results}

For each of the 16 target stars listed in Table \ref{targetList}
we simulate 1000 random orbits and then extract orbital parameters as described above.
Table~\ref{simParamTable} lists the number of orbital calculations for the full set of simulations (see \cite{foreman-mackey13} for details on MCMC parameters).

\begin{deluxetable}{l|c}
\tablecaption{Simulation Parameters
\label{simParamTable}}
\tablehead{Parameter & Quantity }
\startdata
Target stars & 16 \\
Random orbits per star & 1000 \\
MCMC iterations per orbit & 5000 \\
MCMC walkers & 100 \\
\hline
\enddata
\end{deluxetable}

Our first objective is to accurately determine each planet's semi-major axis. The ability to make this measurement depends not only on the astrometric precision for individual observations (2.8 to 5.9 mas; Table~\ref{targetList}), but also depends critically on the orbital sampling. If planets do not trace out their full orbit during the two-year observing window,
the quality of the fit is reduced. This is particularly true for stars with high luminosity, which translates to a more distant habitable zone and hence longer orbital periods. Sirius, the most luminous star in our sample (30.5 $\LSun$), has the worst precision in its orbit fitting (72.5 mas), whereas eps Ind, the least luminous star (0.23 $\LSun$), has the best determined orbit (3.5 mas). (Table \ref{targetList} lists the median semi-major axis precision obtained for the other target stars.)

Our ultimate objective is to determine whether a planet lies within its parent star's habitable zone.  The key metric for this determination is not the absolute precision, but rather the fractional precision on a planet's semi-major axis. While the absolute precision varies between Sirius and eps Ind by a factor of 20, the fractional precision for the two is comparable (2.7\% and 2.1\%, respectively), since Sirius' habitable zone is a factor of 16 larger than eps Ind's. For the overall sample, the median fractional precision varies from 2.1\% up to 7.6\% for Fomalhaut.

\begin{figure}\begin{center}
    \includegraphics[width=2.3in]{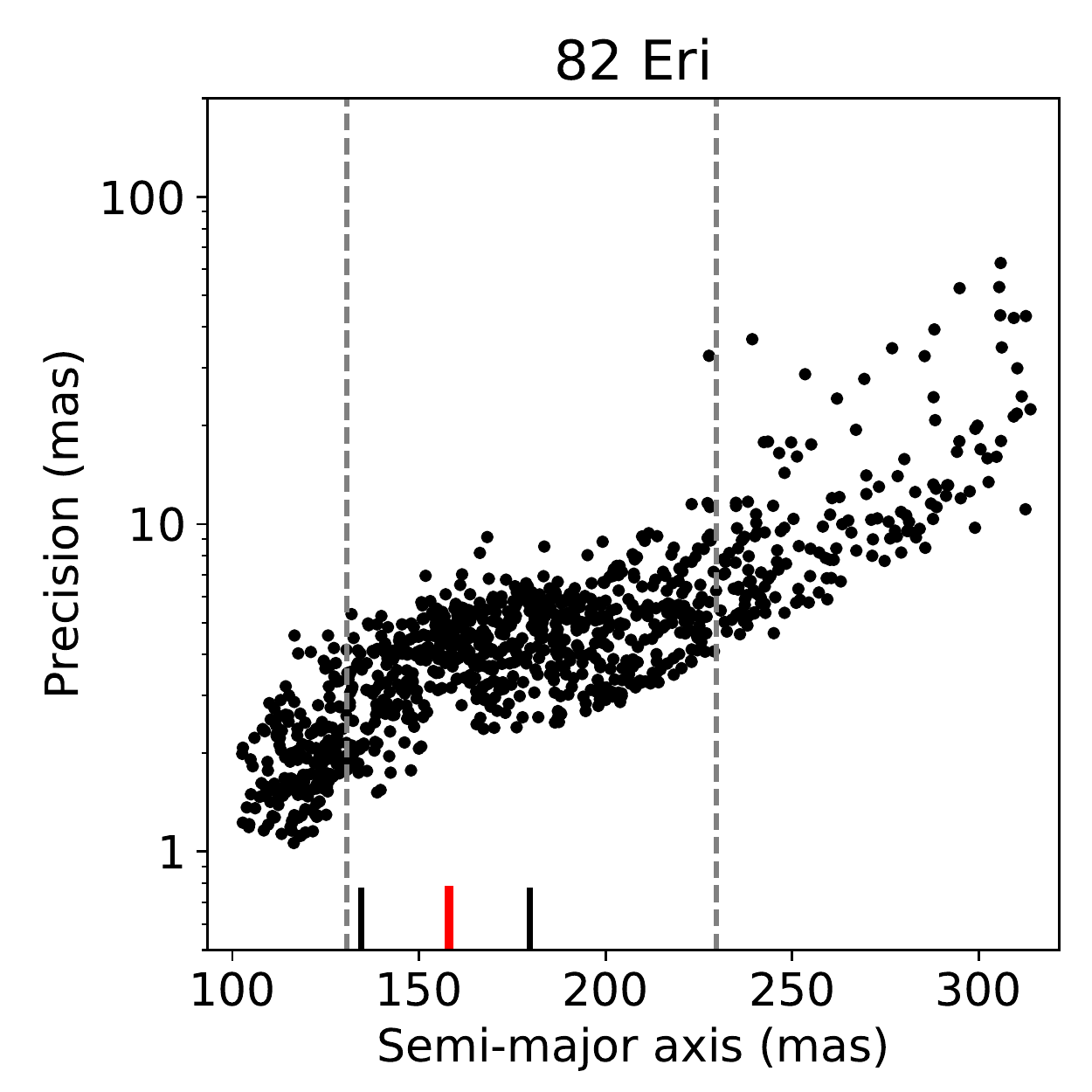}
    \includegraphics[width=2.3in]{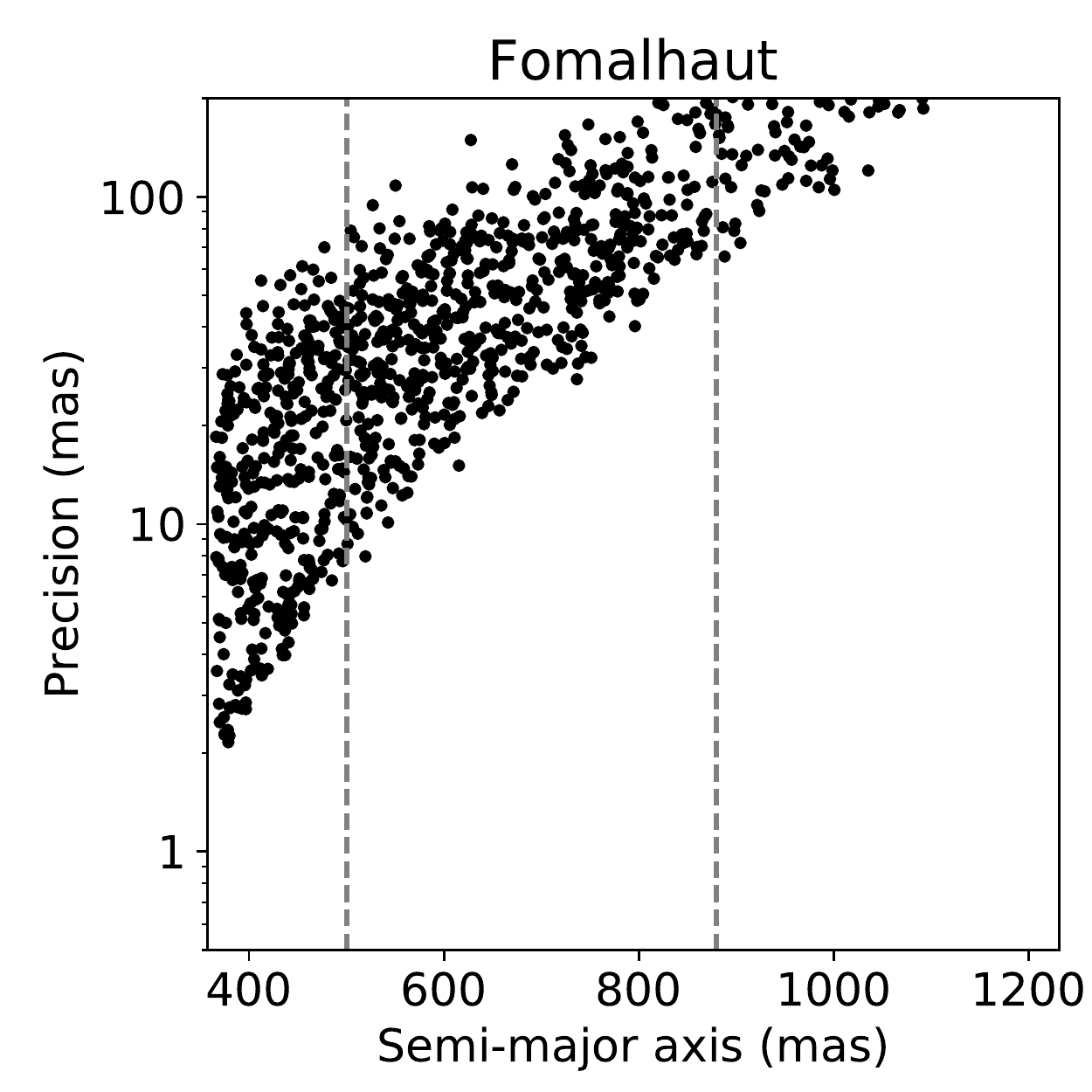}
    \includegraphics[width=2.3in]{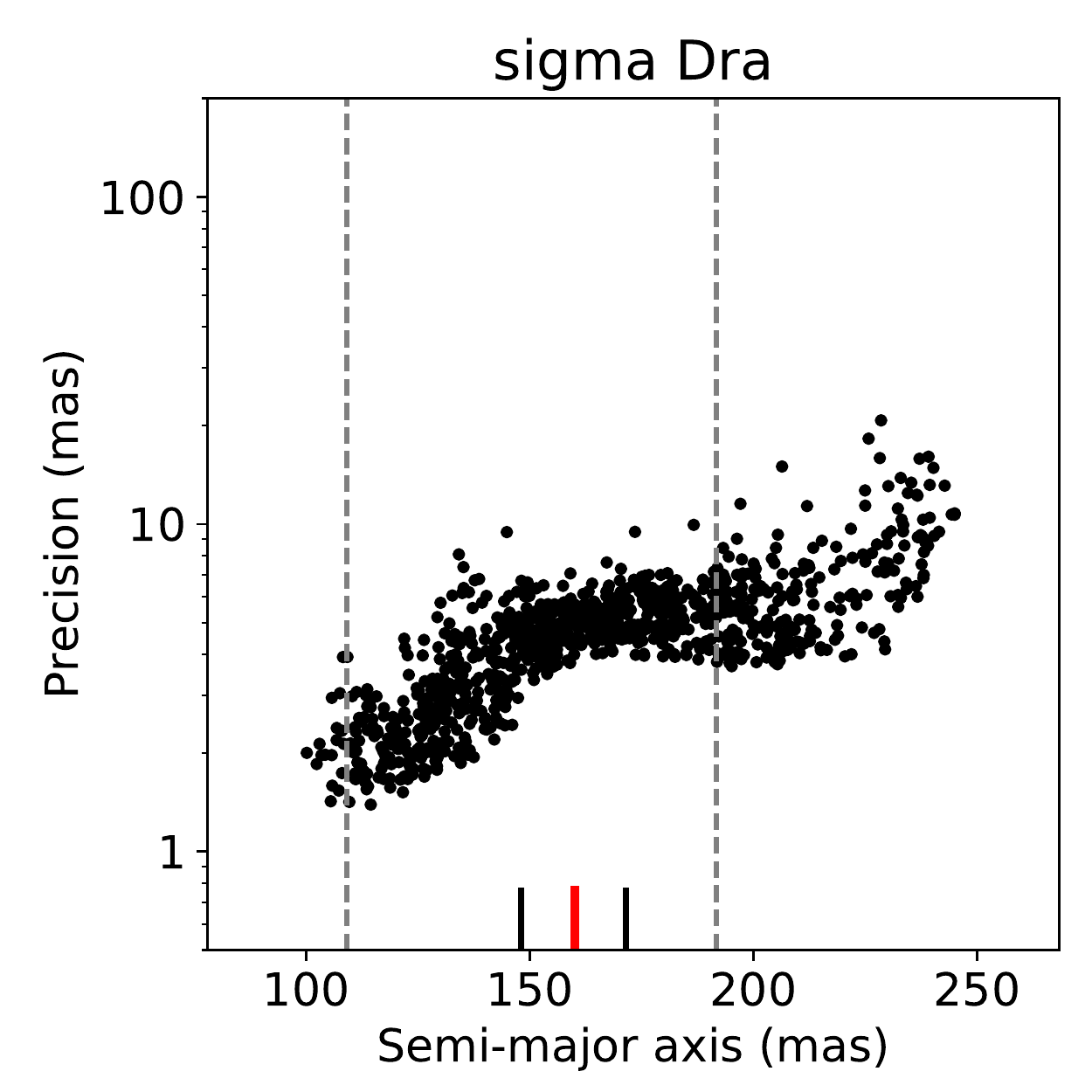}
  \end{center}\caption{
For each star (three examples shown here -- 82 Eri, Fomalhaut, and sigma Dra), the orbital parameters and their uncertainties are retrieved for 1000 random planet orbits, each of which is directly imaged at least three times. The precision for measuring the semi-major axis of each planet is shown here as a function of the true semi-major axis. The habitable zone is interior to the dashed lines. The starshade masks all orbits inside of 100 mas. Hash marks at the bottom of each panel correspond to orbital periods equal to the spacing between observing epochs, with a 1-year period highlighted in red.
  }\label{precisionFig}
\end{figure}

\begin{figure}\begin{center}
    \includegraphics[width=3in]{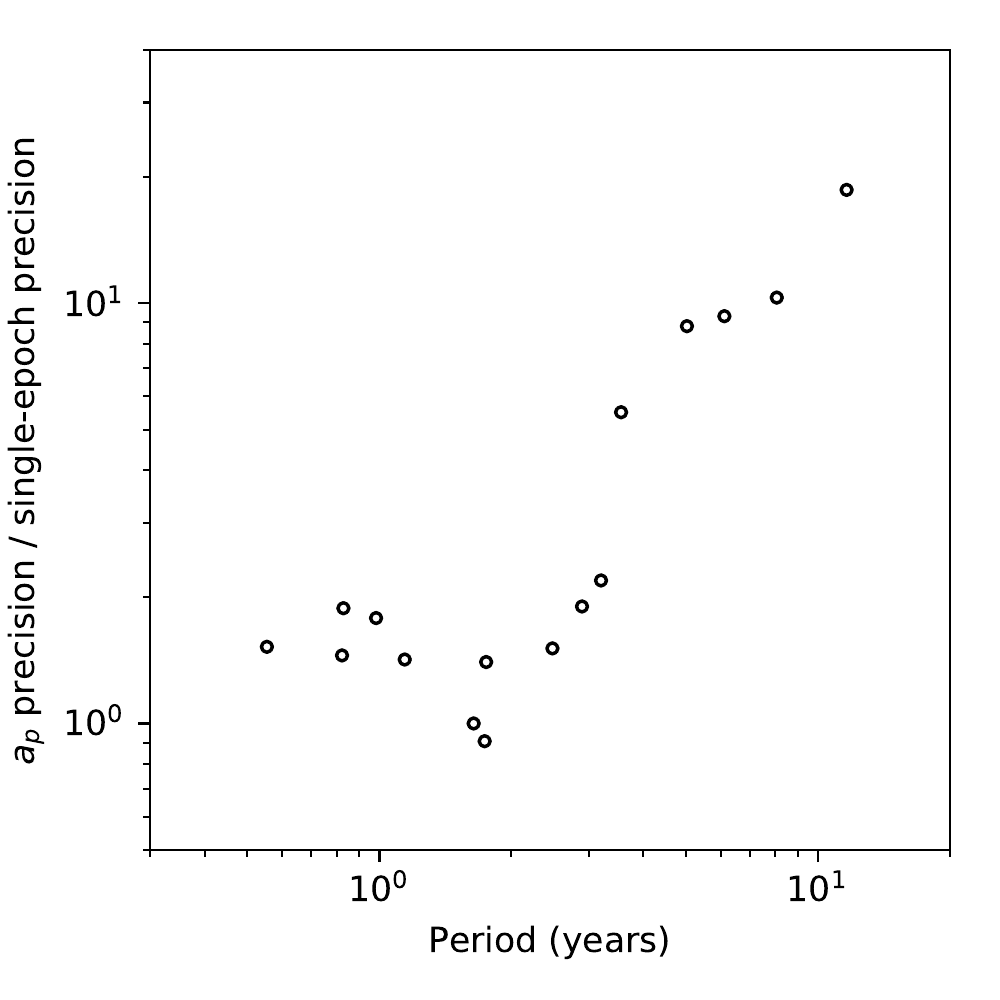}
  \end{center}\caption{
Our ability to pin down a planet's semi-major axis depends on the fraction of its orbit that is traced by the observations. Periods less than the mission lifetime (2 years) are well sampled, while those with longer periods are only observed for a partial arc, resulting in lower precision in determining the orbit.
  }\label{periodDependence}
\end{figure}

Figure~\ref{precisionFig} shows semi-major axis measurement precision versus true semi-major axis for each of the 1000 simulated planets around three of our target stars. 82 Eri (left panel) has one of the best precisions (4.4 mas median), although the performance degrades significantly for more distant orbits, where the planets are relatively faint. Fomalhaut (central panel) has one of the worst precisions (50.5 mas median), primarily because the planets in its habitable zone around this A-type star have periods considerably longer than our 2-year mission lifetime (5--12 years), such that only a fraction of each orbit is traced. The effect of limited phase coverage can be seen in Figure~\ref{periodDependence}, which plots semi-major axis precision as a function of orbital period in the center of the habitable zone.  Planets with periods less than our 2-year mission lifetime are well constrained, but those farther out have semi-major axis precision increasing roughly linearly with the period.

While the semi-major precisions for other target stars (shown in Figure \ref{extraPrecisions} in the Appendix) follow a similar pattern of smoothly decreasing precision with increasing $a_p$, sigma Dra (right panel in Figure~\ref{precisionFig}) has an unusual bump around $a_p \simeq $160 mas, corresponding to orbital periods of $\sim$1 year. The decrease in performance is due to the sampling being repeated on 1-year cycles, where observations taken during the second year of the mission have about the same orbital anomaly as those taken in the first year of the mission (i.e. there is a 1-year aliasing). The resulting small range of orbital-anomaly coverage (see the right panel of Figure \ref{orbitFit}) makes it more difficult to fit the orbit.  

While the 1-year aliases just discussed is most pronounced for sigma Dra, several other systems exhibit a similar effect at orbital periods that match the phasing of the observations. For all of the precision plots (Figures~\ref{precisionFig} and \ref{extraPrecisions}), the semi-major axis corresponding to a 1-year period is indicated by a red hash mark; the other black marks correspond to other time differences between observing epochs (e.g. for 82 Eri, the first and second observations are separated by 40 days, the second and third by 325 days, and the first and fourth by 405 days; see the observing windows in Figure \ref{habitability_windows}). While the 1-year aliasing generally causes the strongest effect, other orbital period/observing period alignments can also degrade performance.

\begin{figure}\begin{center}
    \includegraphics[width=2.3in]{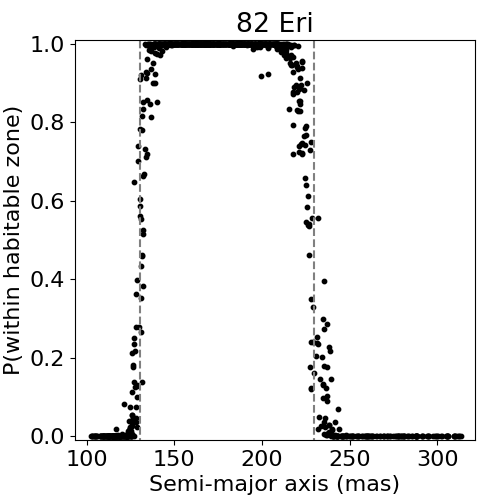}
    \includegraphics[width=2.3in]{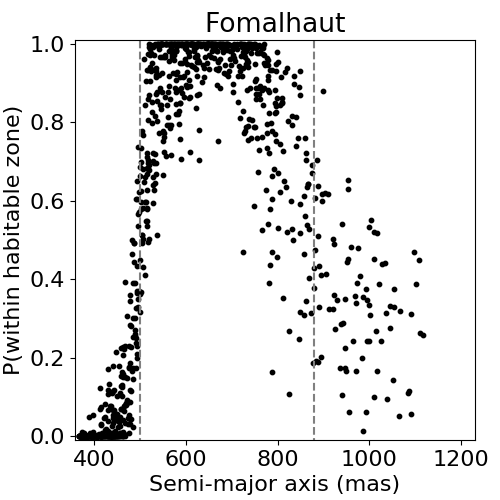}
    \includegraphics[width=2.3in]{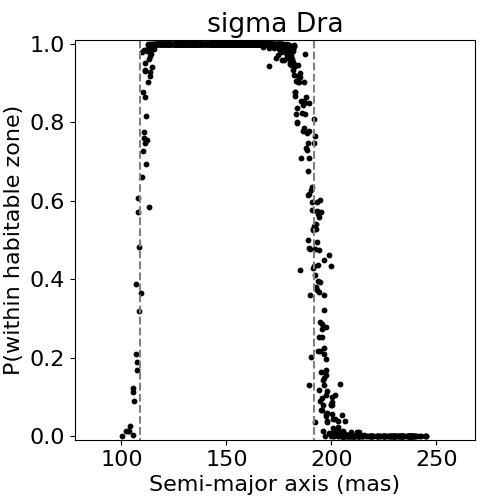}
  \end{center}\caption{
As in Figure \ref{precisionFig}, orbital parameters are measured for 1000 random planetary orbits around each of three stars --
82 Eri, Fomalhaut, and sigma Dra. The derived probability of each planet residing in its star's habitable zone is is shown as a function of its true semi-major axis. The habitable zone is indicated by the dashed lines.
  }\label{HZprobFig}
\end{figure}

Our MCMC fitting procedure calculates a (non-Gaussian) posterior distribution for each orbital parameter.  From these distributions we derive the probability that each planet lies within its star's habitable zone. Figure~\ref{HZprobFig} shows the results for the same three target stars as in Figure \ref{precisionFig}. For 82 Eri (left panel), the orbit fitting is fairly deterministic -- planets well inside the habitable zone are correctly identified as such with high probability ($>$99\%), while those well outside are ruled out (probability $<$1\%).  As one would expect, there is some ambiguity near the edges of the habitable zone, but for the overall sample there is just a 2.4\% chance of a habitable zone planet being falsely classified as  falling outside the habitable zone, while 98.8\% of the planets classified as residing in the habitable zone are truly habitable zone planets (i.e.\ a false positive rate of 1.2\%).

\begin{figure}\begin{center}
     \includegraphics[width=2.3in]{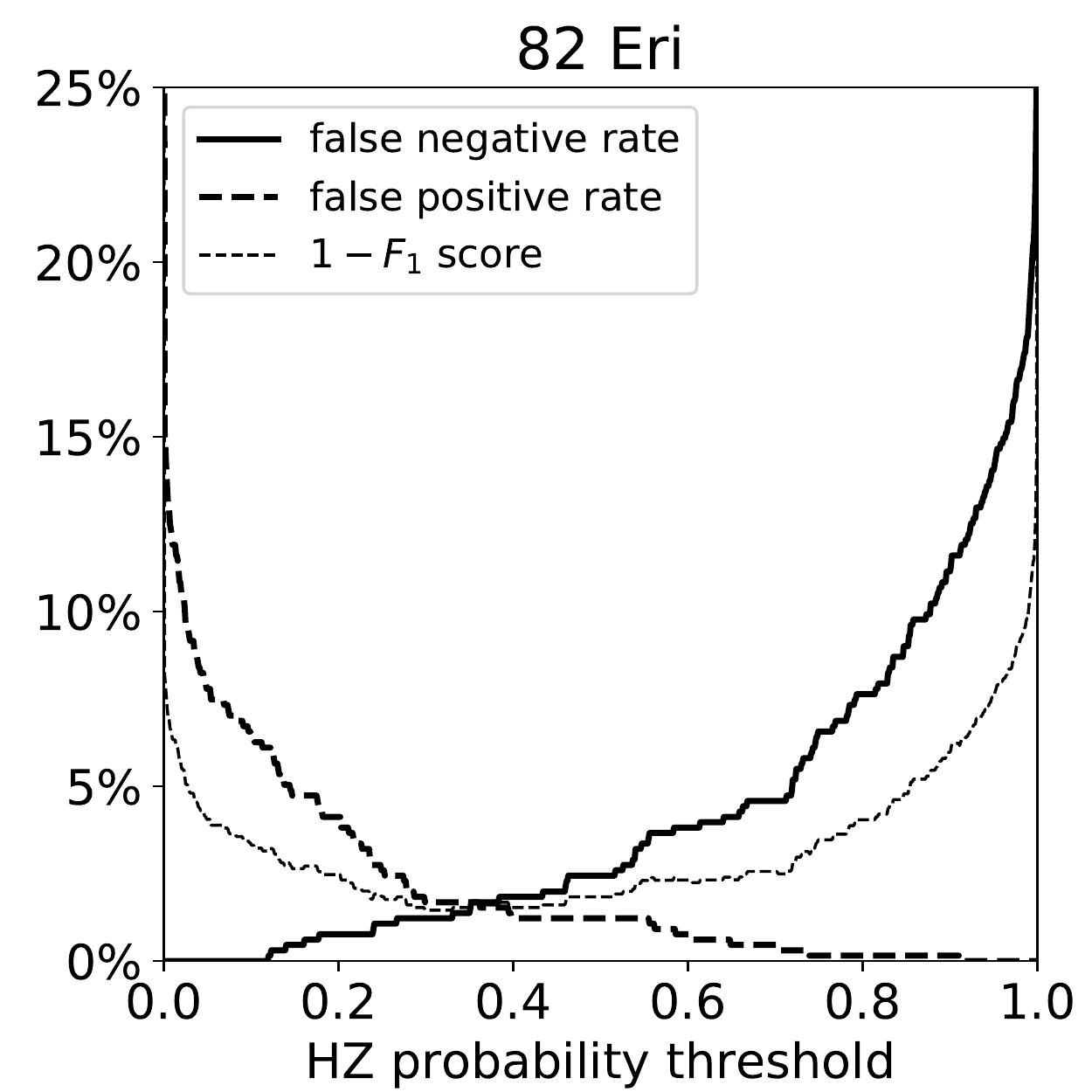}
     \includegraphics[width=2.3in]{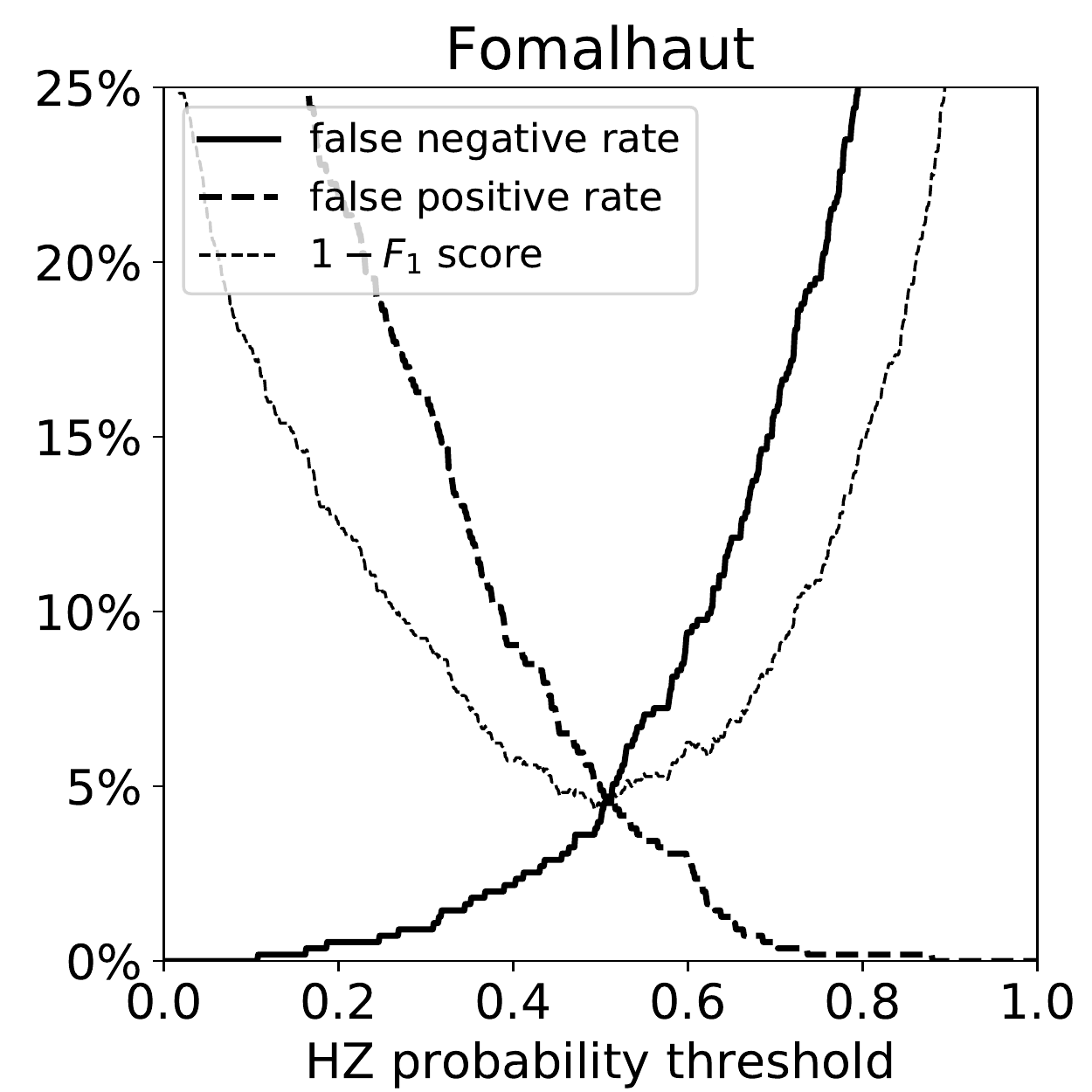}
     \includegraphics[width=2.3in]{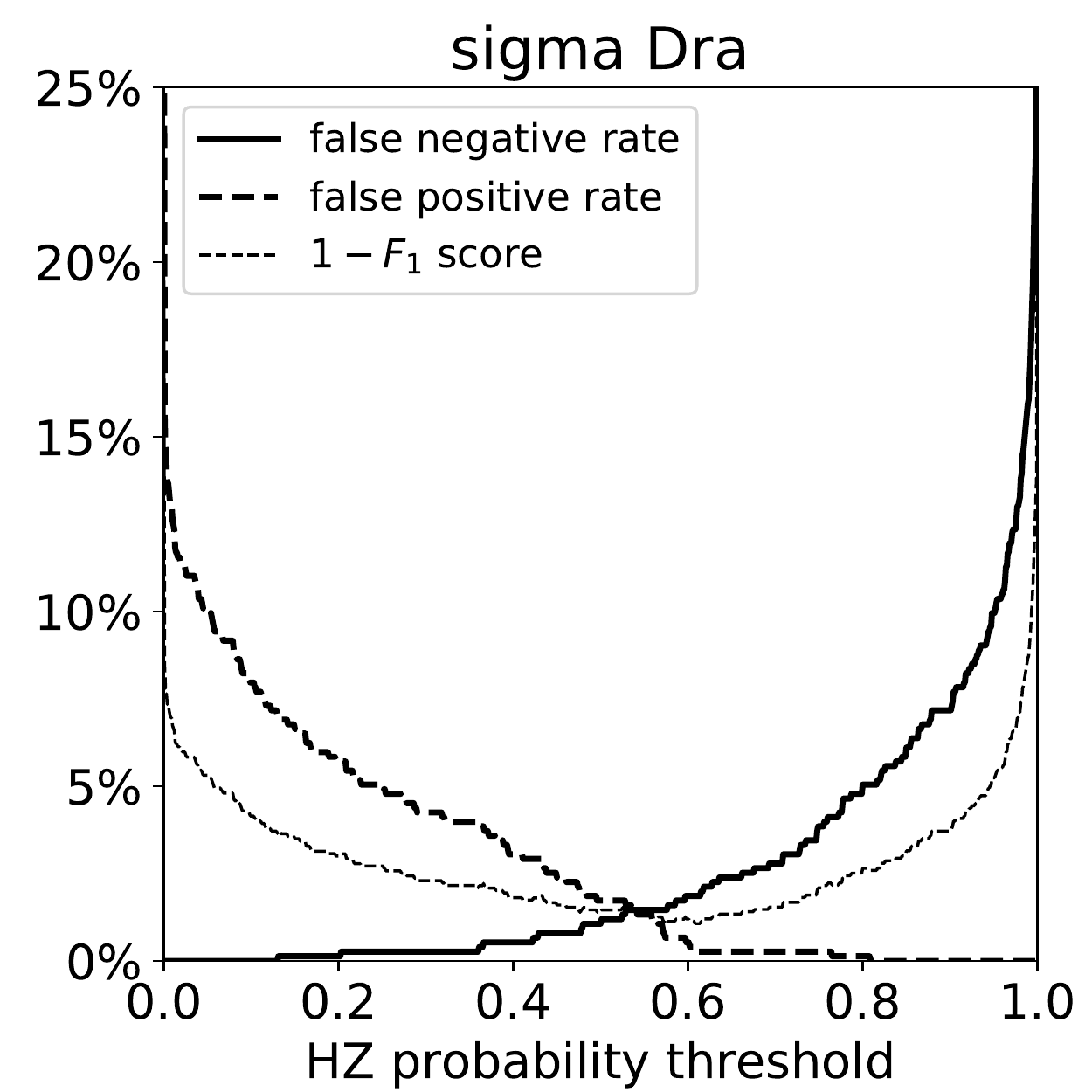}
  \end{center}
  \caption{
    False  positive and false negative rates for 82 Eri, Fomalhaut, and sigma Dra, shown as a function of the habitable-zone probability threshold. A more liberal threshold (lower probability) reduces the false negative rate, but increases the false positive rate. A very conservative threshold (to the right of each panel) can ensure that no false detections are made, but misses a significant number of true habitable-zone planets. A performance metric combining the two rates is given by the $F_1$ score, the harmonic mean of the precision (the fraction of detected habitable zone planets that truly are located in the habitable zone) and recall (the fraction of habitable zone planets that are correctly classified as such).
  }\label{falseratesFig}
\end{figure}

These rates are based on a nominal classification threshold, where planets with habitable zone probability $>$50\% are categorized as habitable zone planets. If a more conservative approach is desired, less planets can be included 
as habitable zone. If a 95\% threshold is used for classification, for example, then 82 Eri will have only 0.2\% false detections.  However, only 88.9\% of the true habitable zone planets will be included (11.1\% false negative rate).

For Fomalhaut (middle panel of Figure \ref{HZprobFig}), the worse orbital precision translates to much more scatter in the plotted probabilities and less certainty for determining whether a planet lies in its habitable zone. Still, there is only 4.0\% probability of a habitable zone planet being misclassified, and only a 4.9\% chance of a habitable-zone-classified planet not being truly in the habitable zone. For the conservative (95\%) classification threshold, the false positive rate falls to zero, but at the expense of only 59\% of the true habitable zone planets being included (i.e.\ a false negative rate of 41\%).

The false positive and false negative rates for 82 Eri, Fomalhaut, and sigma Dra are shown in Figure \ref{falseratesFig}, as a function of the classification threshold. These plots also display an overall success metric -- the $F_1$ score,  the harmonic mean of (1 - false positive rate) and (1 - false negative rate). While a nominal threshold of 50\% results in an optimal balance between these two factors (i.e.\ it give maximum $F_1$ score), an emphasis on avoiding false detections would warrant a more conservative approach. 

The false positive/false negative rates for each star are listed in Table \ref{targetList} for both the nominal classification threshold (50\%) and a conservative classification threshold (95\%). For the nominal threshold, the average performance for the overall sample is a 2.8\% false positive rate and a 3.3\% false negative rate. For the conservative threshold, the average false positive rate is just 0.05\%, but the average false negative rate goes up to 19\%.


\section{Conclusions}\label{summary}

Based  on a model for the Starshade Rendezvous Probe (SRP) mission concept with target-specific observing windows and SNR calculations dependent on the planet illumination during each window, we have quantified the ability of SRP to identify habitable zone planets. We find that detection of a planet in at least 3 out of the 4 observing epochs will adequately measure the planet's semi-major axis. For a 16 star sample observed with this strategy, we find that habitable zone planets are correctly identified as such 96.7\% of the time, with 2.8\% contamination by false classifications. Including the full range of planet masses, the mission is expected to detect $\sim$10 planets in the vicinity of the habitable zone \citep{paper1}, such that a very small number of planets (less than 1) are expected to be misclassified.


\acknowledgments
{\it Acknowledgements:}
Part of this work was carried out at the Jet Propulsion Laboratory, California Institute of Technology, under a contract with the National Aeronautics and Space Administration. \copyright 2020. All rights reserved. This research has made use of 1) the NASA Exoplanet Archive, which is operated by the California Institute of Technology, under contract with the National Aeronautics and Space Administration under the Exoplanet Exploration Program and 2) the SIMBAD database, operated at CDS, Strasbourg, France. 


\footnotesize
\bibliographystyle{aasjournal}
\bibliography{refs}

\begin{thebibliography}{}
\expandafter\ifx\csname natexlab\endcsname\relax\def\natexlab#1{#1}\fi
\providecommand{\url}[1]{\href{#1}{#1}}
\providecommand{\dodoi}[1]{doi:~\href{http://doi.org/#1}{\nolinkurl{#1}}}
\providecommand{\doeprint}[1]{\href{http://ascl.net/#1}{\nolinkurl{http://ascl.net/#1}}}
\providecommand{\doarXiv}[1]{\href{https://arxiv.org/abs/#1}{\nolinkurl{https://arxiv.org/abs/#1}}}

\bibitem[{{Belikov} \& {et al.}(2017)}]{belikov17}
{Belikov}, R., \& {et al.} 2017,
  https://exoplanets.nasa.gov/exep/exopag/sag/\#sag13

\bibitem[{{Blunt} {et~al.}(2017){Blunt}, {Nielsen}, {De Rosa}, {Konopacky},
  {Ryan}, {Wang}, {Pueyo}, {Rameau}, {Marois}, {Marchis}, {Macintosh},
  {Graham}, {Duch{\^e}ne}, \& {Schneider}}]{blunt17}
{Blunt}, S., {Nielsen}, E.~L., {De Rosa}, R.~J., {et~al.} 2017, \aj, 153, 229,
  \dodoi{10.3847/1538-3881/aa6930}

\bibitem[{{Ertel} {et~al.}(2020){Ertel}, {Defr{\`e}re}, {Hinz}, {Mennesson},
  {Kennedy}, {Danchi}, {Gelino}, {Hill}, {Hoffmann}, {Mazoyer}, {Rieke},
  {Shannon}, {Stapelfeldt}, {Spalding}, {Stone}, {Vaz}, {Weinberger},
  {Willems}, {Absil}, {Arbo}, {Bailey}, {Beichman}, {Bryden}, {Downey},
  {Durney}, {Esposito}, {Gaspar}, {Grenz}, {Haniff}, {Leisenring}, {Marion},
  {McMahon}, {Millan-Gabet}, {Montoya}, {Morzinski}, {Perera}, {Pinna}, {Pott},
  {Power}, {Puglisi}, {Roberge}, {Serabyn}, {Skemer}, {Su}, {Vaitheeswaran}, \&
  {Wyatt}}]{ertel20}
{Ertel}, S., {Defr{\`e}re}, D., {Hinz}, P., {et~al.} 2020, \aj, 159, 177,
  \dodoi{10.3847/1538-3881/ab7817}

\bibitem[{{Ford}(2006)}]{ford06}
{Ford}, E.~B. 2006, \pasp, 118, 364, \dodoi{10.1086/500813}

\bibitem[{{Foreman-Mackey} {et~al.}(2013){Foreman-Mackey}, {Hogg}, {Lang}, \&
  {Goodman}}]{foreman-mackey13}
{Foreman-Mackey}, D., {Hogg}, D.~W., {Lang}, D., \& {Goodman}, J. 2013, \pasp,
  125, 306, \dodoi{10.1086/670067}

\bibitem[{{Guimond} \& {Cowan}(2019)}]{guimond19}
{Guimond}, C.~M., \& {Cowan}, N.~B. 2019, \aj, 157, 188,
  \dodoi{10.3847/1538-3881/ab0f2e}

\bibitem[{{Guyon} {et~al.}(2013){Guyon}, {Eisner}, {Angel}, {Woolf}, {Bendek},
  {Milster}, {Ammons}, {Shao}, {Shaklan}, {Levine}, {Nemati}, {Martinache},
  {Pitman}, {Woodruff}, \& {Belikov}}]{guyon13}
{Guyon}, O., {Eisner}, J.~A., {Angel}, R., {et~al.} 2013, \apj, 767, 11,
  \dodoi{10.1088/0004-637X/767/1/11}

\bibitem[{{Horning} {et~al.}(2019){Horning}, {Morgan}, \&
  {Nielson}}]{horning19}
{Horning}, A., {Morgan}, R., \& {Nielson}, E. 2019, in Society of Photo-Optical
  Instrumentation Engineers (SPIE) Conference Series, Vol. 11117, Society of
  Photo-Optical Instrumentation Engineers (SPIE) Conference Series, 111171C,
  \dodoi{10.1117/12.2529741}

\bibitem[{{Kasting} {et~al.}(1993){Kasting}, {Whitmire}, \&
  {Reynolds}}]{kasting93}
{Kasting}, J.~F., {Whitmire}, D.~P., \& {Reynolds}, R.~T. 1993, \icarus, 101,
  108, \dodoi{10.1006/icar.1993.1010}

\bibitem[{{Leinert} {et~al.}(1998){Leinert}, {Bowyer}, {Haikala}, {Hanner},
  {Hauser}, {Levasseur-Regourd}, {Mann}, {Mattila}, {Reach}, {Schlosser},
  {Staude}, {Toller}, {Weiland}, {Weinberg}, \& {Witt}}]{leinert98}
{Leinert}, C., {Bowyer}, S., {Haikala}, L.~K., {et~al.} 1998, \aaps, 127, 1,
  \dodoi{10.1051/aas:1998105}

\bibitem[{{Mawet} {et~al.}(2019){Mawet}, {Hirsch}, {Lee}, {Ruffio}, {Bottom},
  {Fulton}, {Absil}, {Beichman}, {Bowler}, {Bryan}, {Choquet}, {Ciardi},
  {Christiaens}, {Defr{\`e}re}, {Gomez Gonzalez}, {Howard}, {Huby}, {Isaacson},
  {Jensen-Clem}, {Kosiarek}, {Marcy}, {Meshkat}, {Petigura}, {Reggiani},
  {Ruane}, {Serabyn}, {Sinukoff}, {Wang}, {Weiss}, \& {Ygouf}}]{mawet19}
{Mawet}, D., {Hirsch}, L., {Lee}, E.~J., {et~al.} 2019, \aj, 157, 33,
  \dodoi{10.3847/1538-3881/aaef8a}

\bibitem[{{Mede} \& {Brandt}(2017)}]{mede17}
{Mede}, K., \& {Brandt}, T.~D. 2017, \aj, 153, 135,
  \dodoi{10.3847/1538-3881/aa5e4a}

\bibitem[{{Robinson} \& {Reinhard}(2018)}]{robinson18}
{Robinson}, T.~D., \& {Reinhard}, C.~T. 2018, arXiv e-prints, 1804,
  arXiv:1804.04138.
\newblock \doarXiv{1804.04138}

\bibitem[{{Rogers}(2015)}]{rogers15}
{Rogers}, L.~A. 2015, \apj, 801, 41, \dodoi{10.1088/0004-637X/801/1/41}

\bibitem[{{Romero-Wolf} {et~al.}(2020){Romero-Wolf}, {Bryden}, {Seager},
  {Somebody}, {Somebody}, {Su}, {Vaitheeswaran}, \& {Wyatt}}]{paper1}
{Romero-Wolf}, A., {Bryden}, G., {Seager}, S., {et~al.} 2020

\bibitem[{{Seager} {et~al.}(2019){Seager}, {Kasdan}, {Romero-Wolf}, {Benneke},
  {Deming}, {Stevenson}, {Seager}, {Berta-Thompson}, {Seifahrt}, \&
  {Homeier}}]{seager19}
{Seager}, S., {Kasdan}, J., {Romero-Wolf}, A., {et~al.} 2019, {Starshade
  Rendezvous Probe},
  https://smd-prod.s3.amazonaws.com/science-red/s3fs-public/atoms/files/Starshade2.pdf

\bibitem[{{Stark} {et~al.}(2016){Stark}, {Shaklan}, {Lisman}, {Cady},
  {Savransky}, {Roberge}, \& {Mand ell}}]{stark16}
{Stark}, C.~C., {Shaklan}, S., {Lisman}, D., {et~al.} 2016, Journal of
  Astronomical Telescopes, Instruments, and Systems, 2, 041204,
  \dodoi{10.1117/1.JATIS.2.4.041204}

\bibitem[{{Turnbull} {et~al.}(2012){Turnbull}, {Glassman}, {Roberge}, {Cash},
  {Noecker}, {Lo}, {Mason}, {Oakley}, \& {Bally}}]{turnbull12}
{Turnbull}, M.~C., {Glassman}, T., {Roberge}, A., {et~al.} 2012, \pasp, 124,
  418, \dodoi{10.1086/666325}

\bibitem[{{Zahnle} \& {Catling}(2017)}]{zahnle17}
{Zahnle}, K.~J., \& {Catling}, D.~C. 2017, \apj, 843, 122,
  \dodoi{10.3847/1538-4357/aa7846}

\end{thebibliography}


\appendix
\section{Additional figures}
\clearpage

\begin{figure}\begin{center}
     \includegraphics[width=2.in]{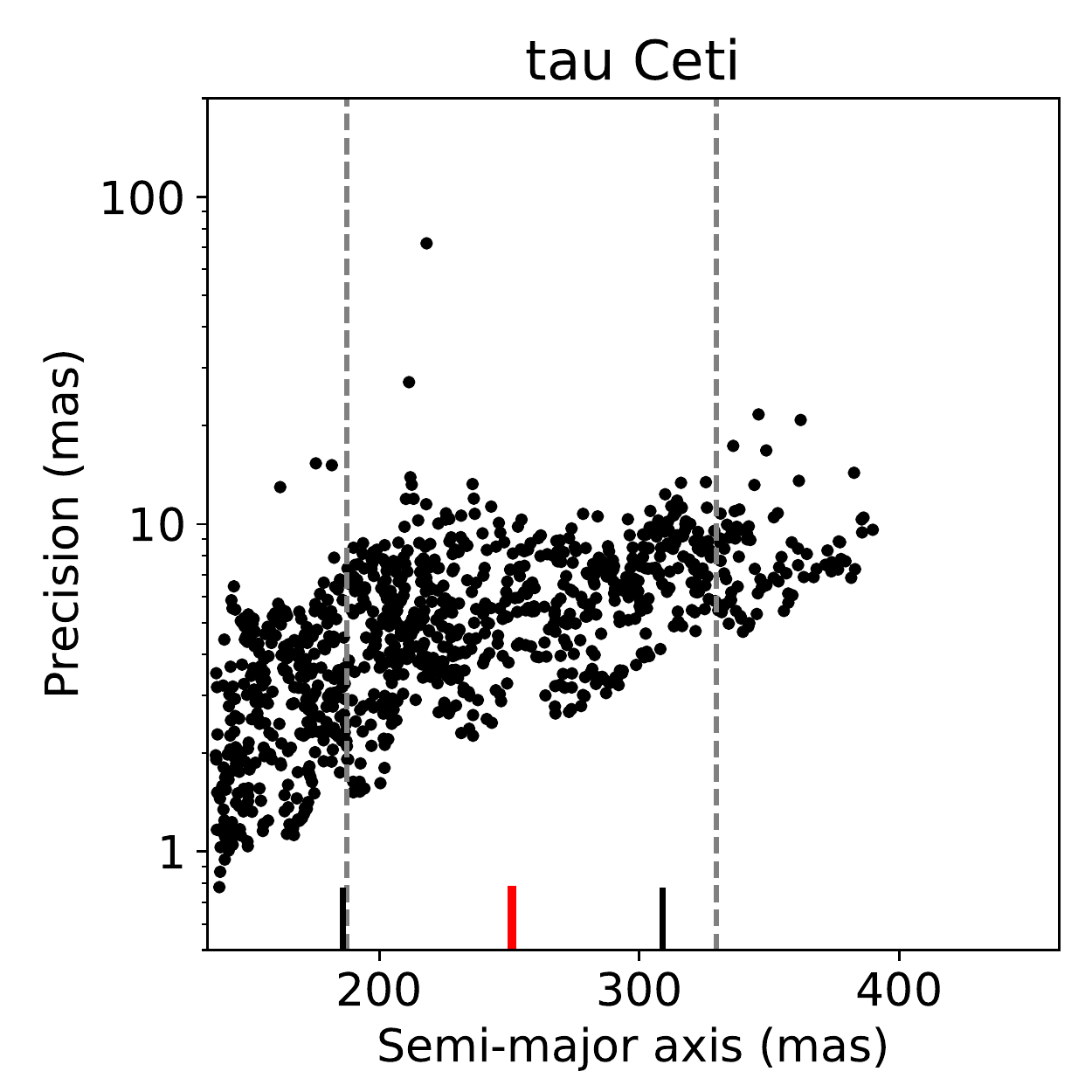}
     \includegraphics[width=2.in]{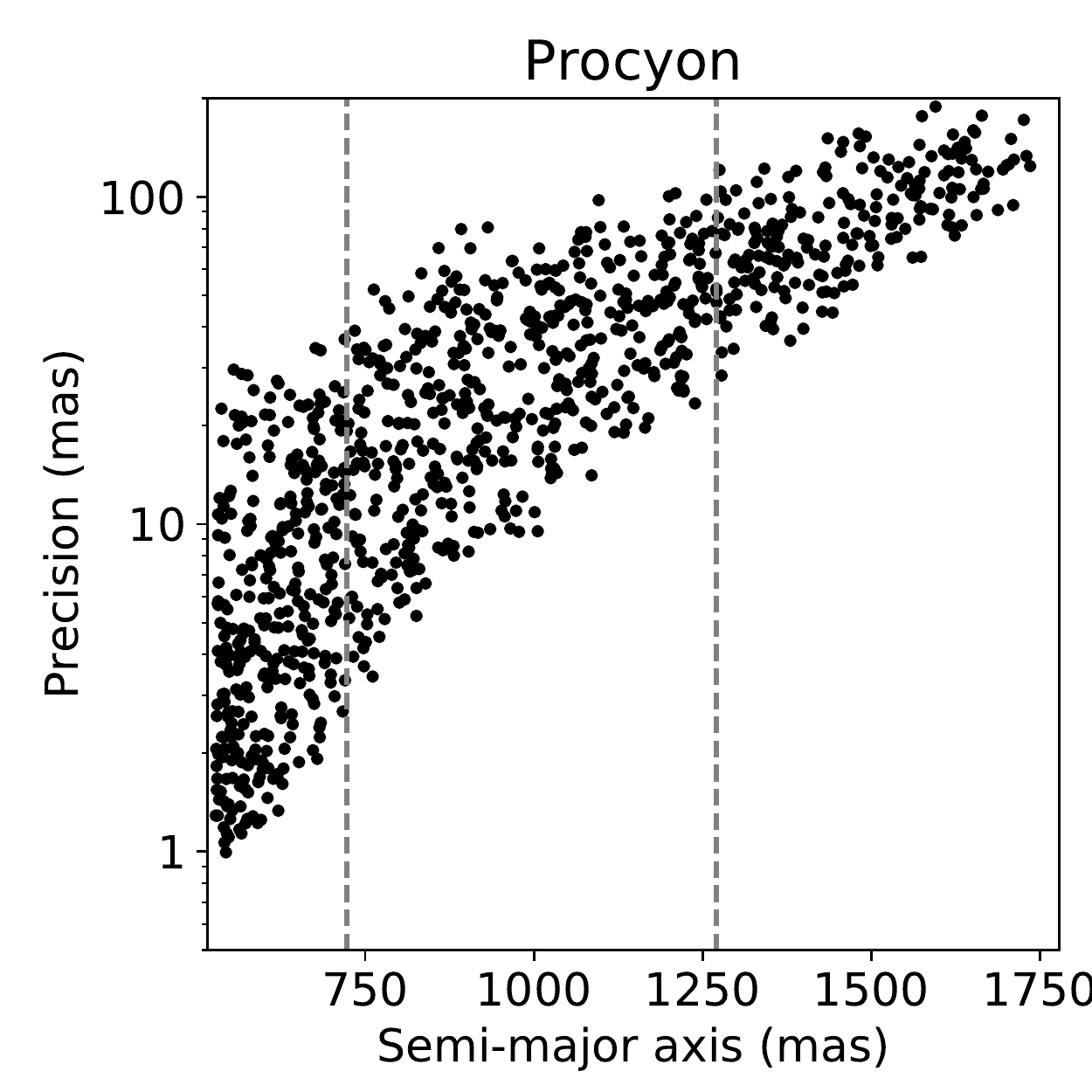}
     \includegraphics[width=2.in]{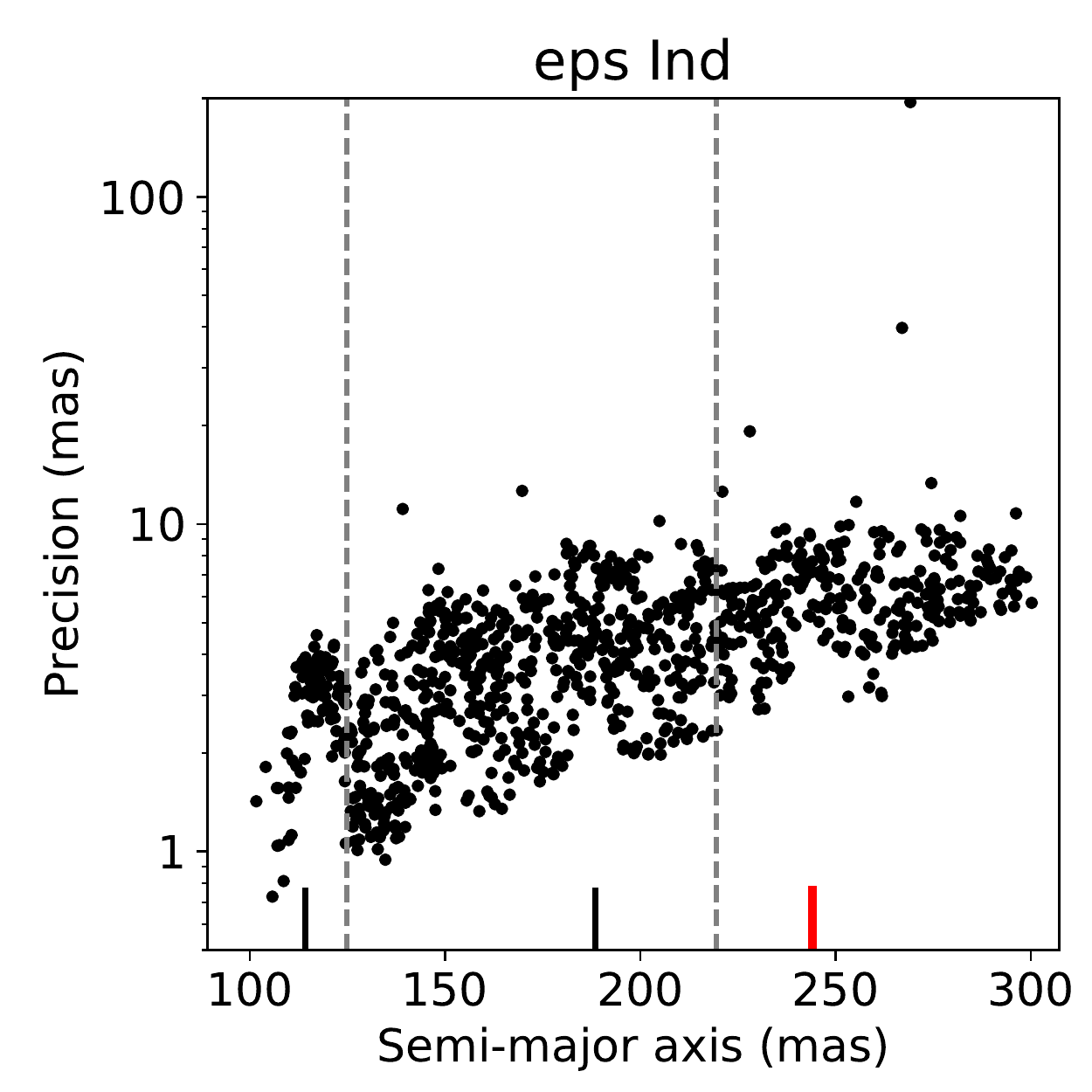}
     \includegraphics[width=2.in]{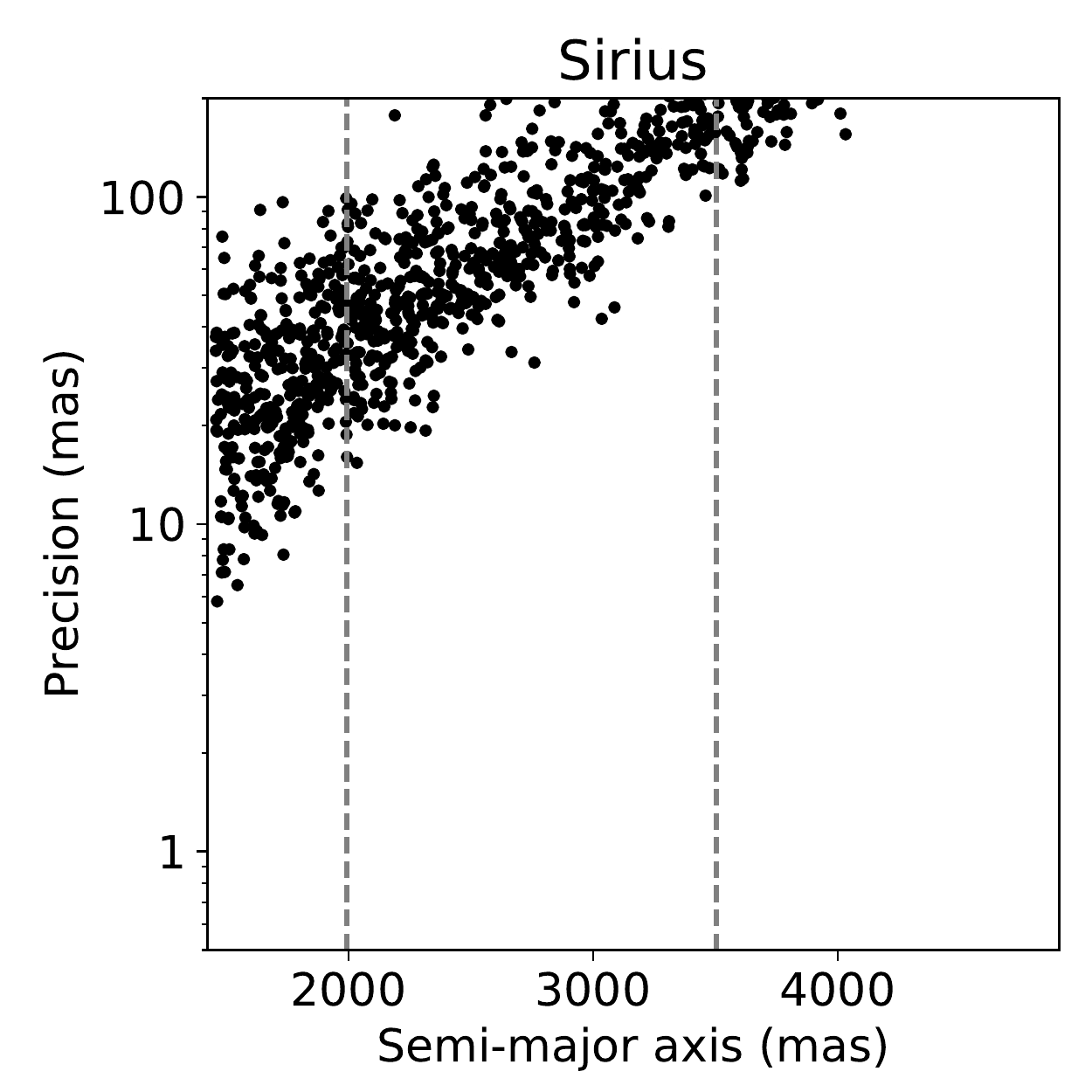}
     \includegraphics[width=2.in]{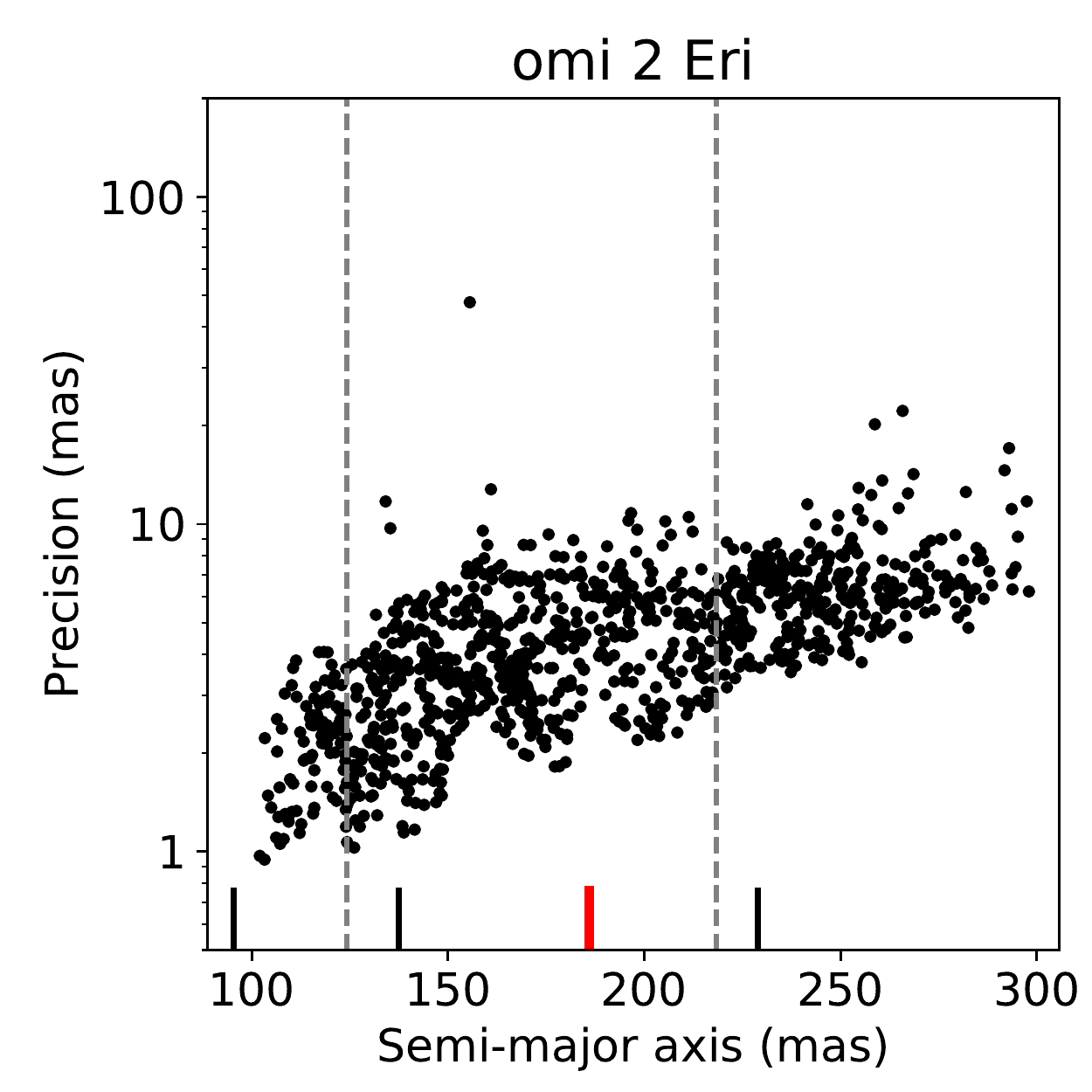}
     \includegraphics[width=2.in]{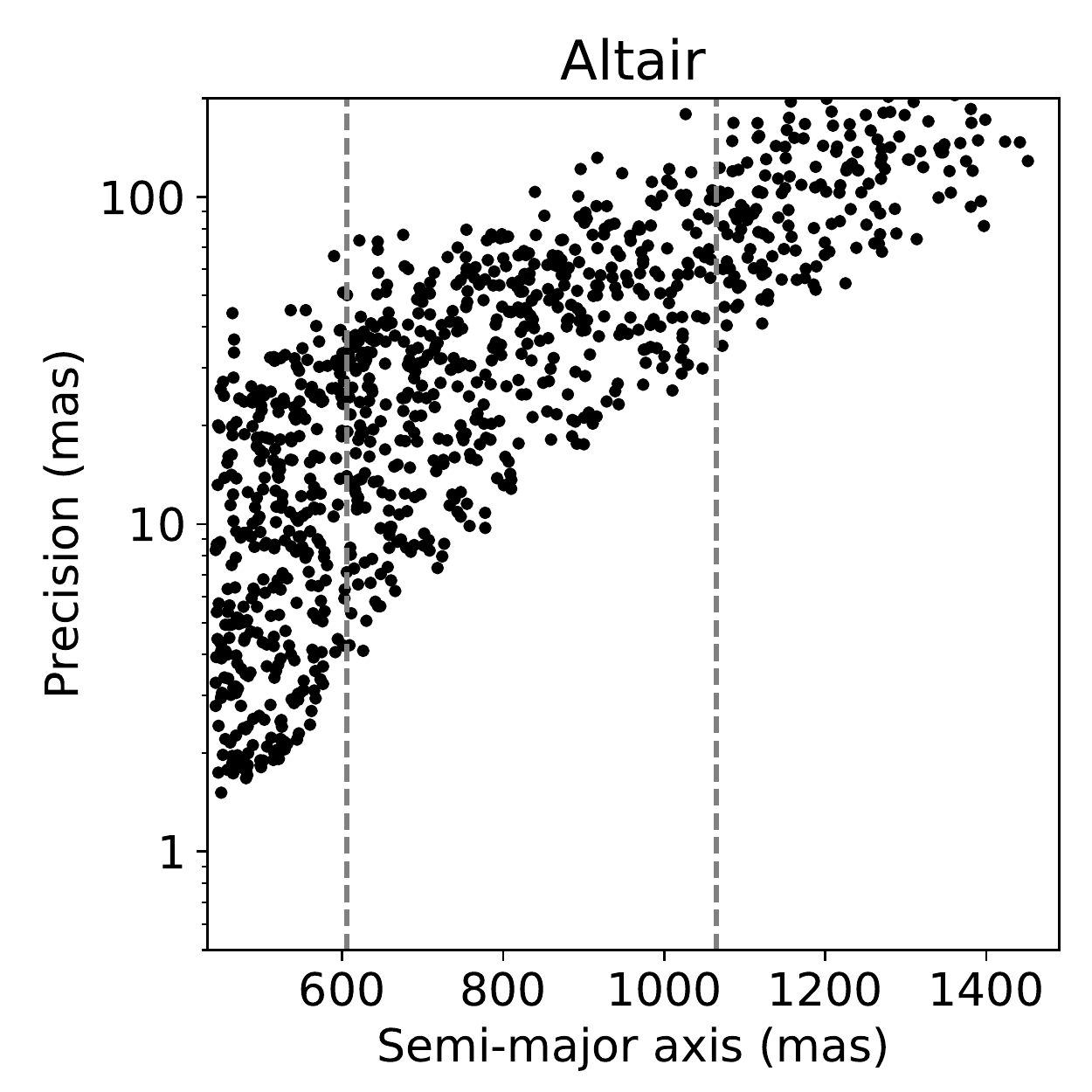}
     \includegraphics[width=2.in]{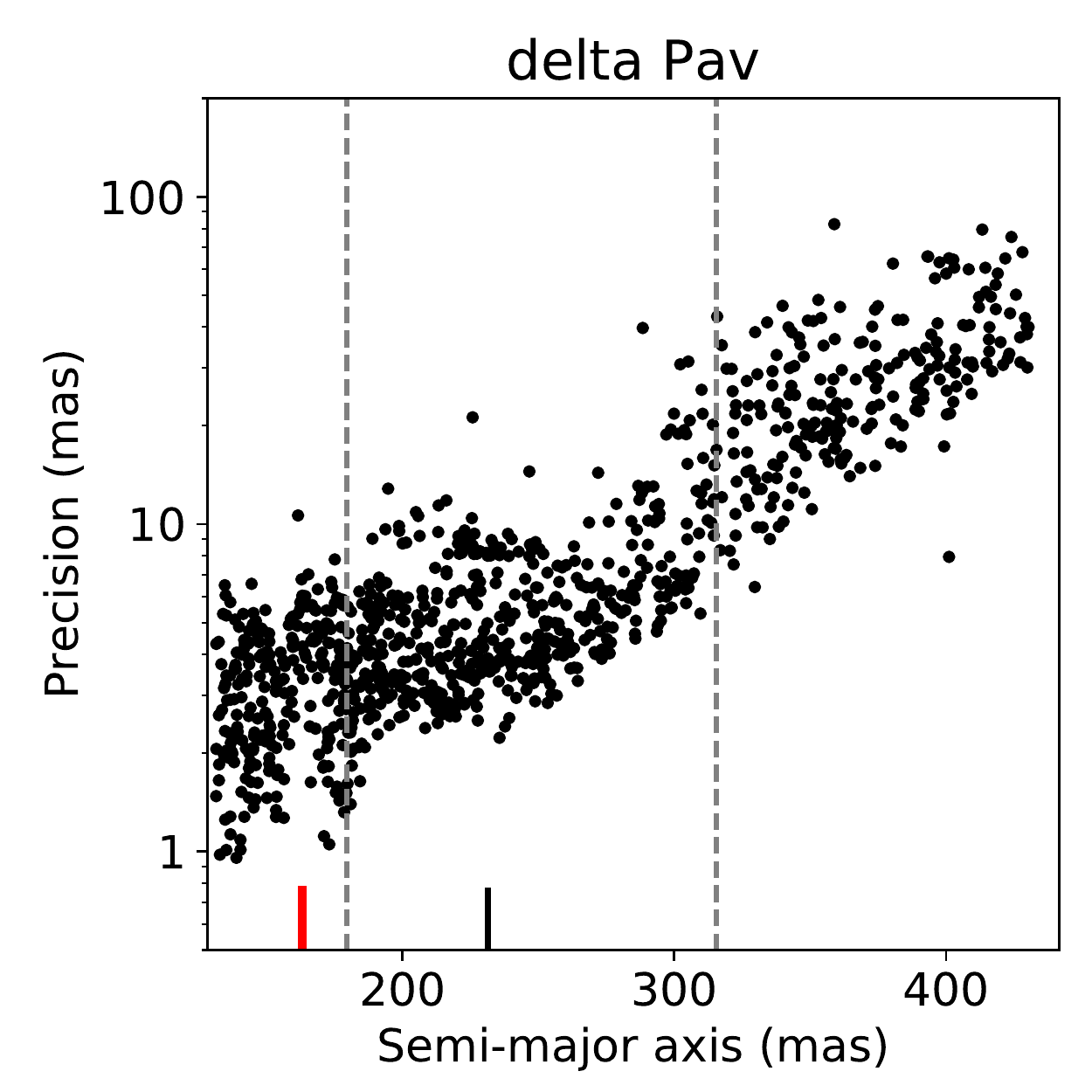}
     \includegraphics[width=2.in]{precision_summary.82Eri.bw.pdf}
     \includegraphics[width=2.in]{precision_summary.sigmaDra.bw.pdf}
  \end{center}\end{figure}\begin{figure}\begin{center}
       \includegraphics[width=2.in]{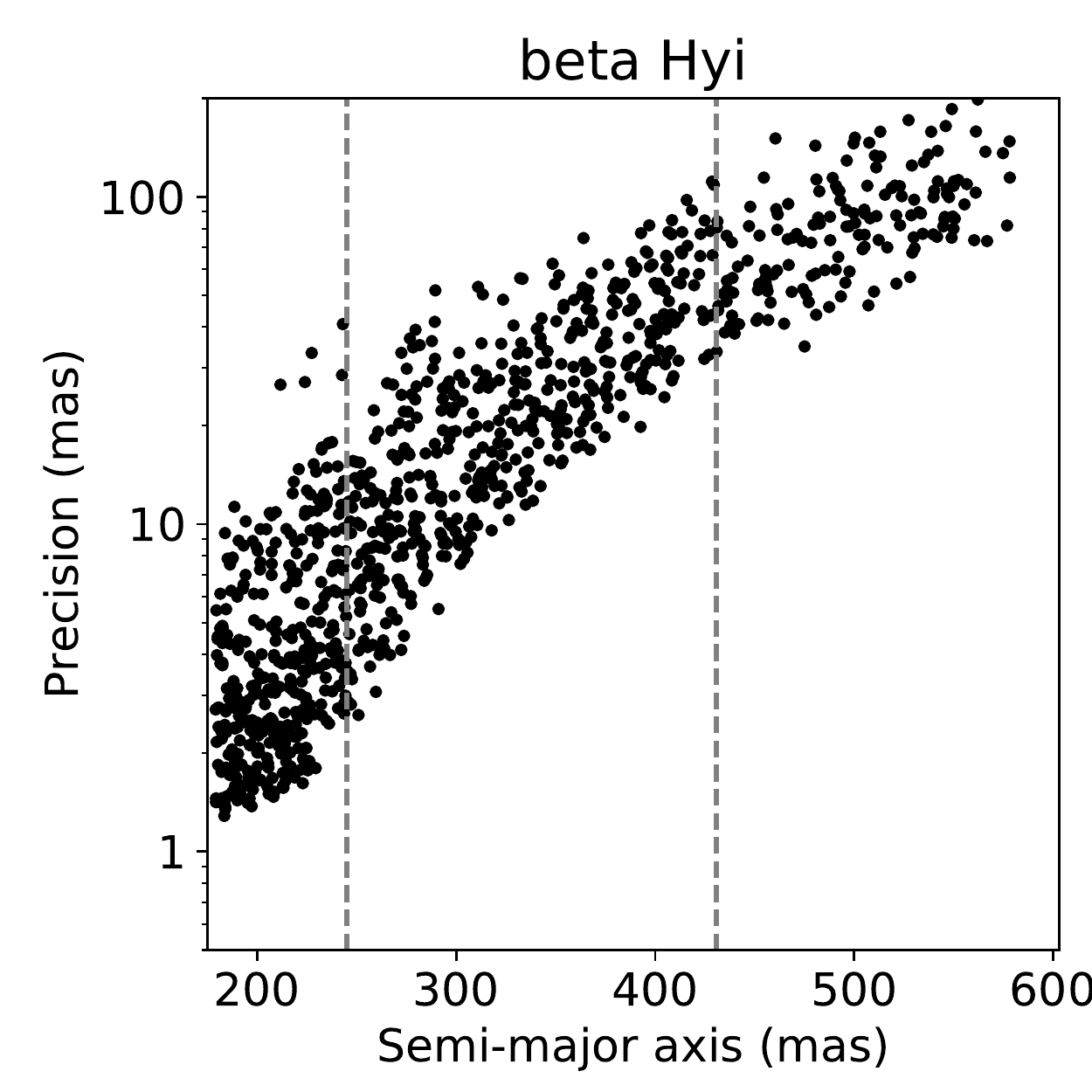}
     \includegraphics[width=2.in]{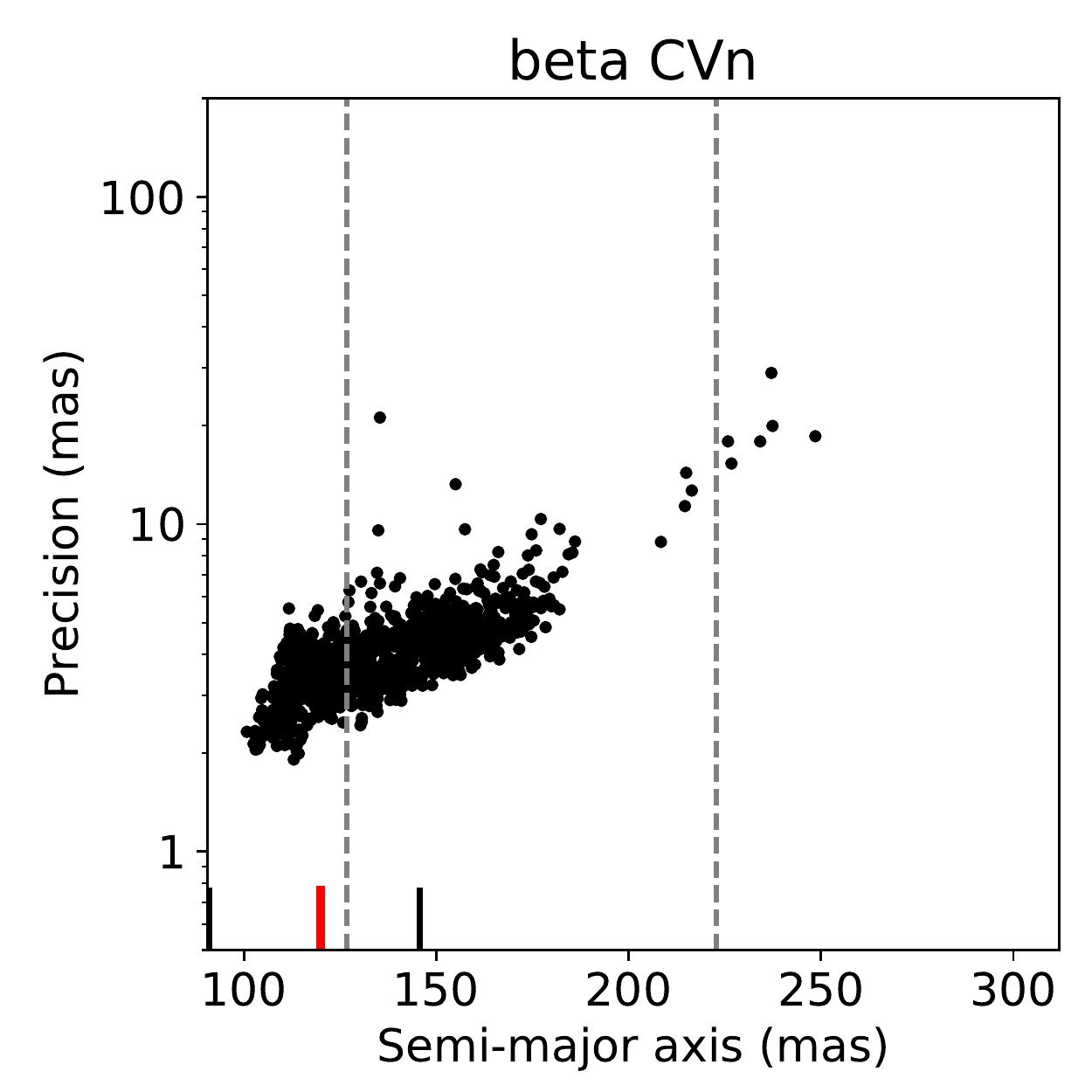}
     \includegraphics[width=2.in]{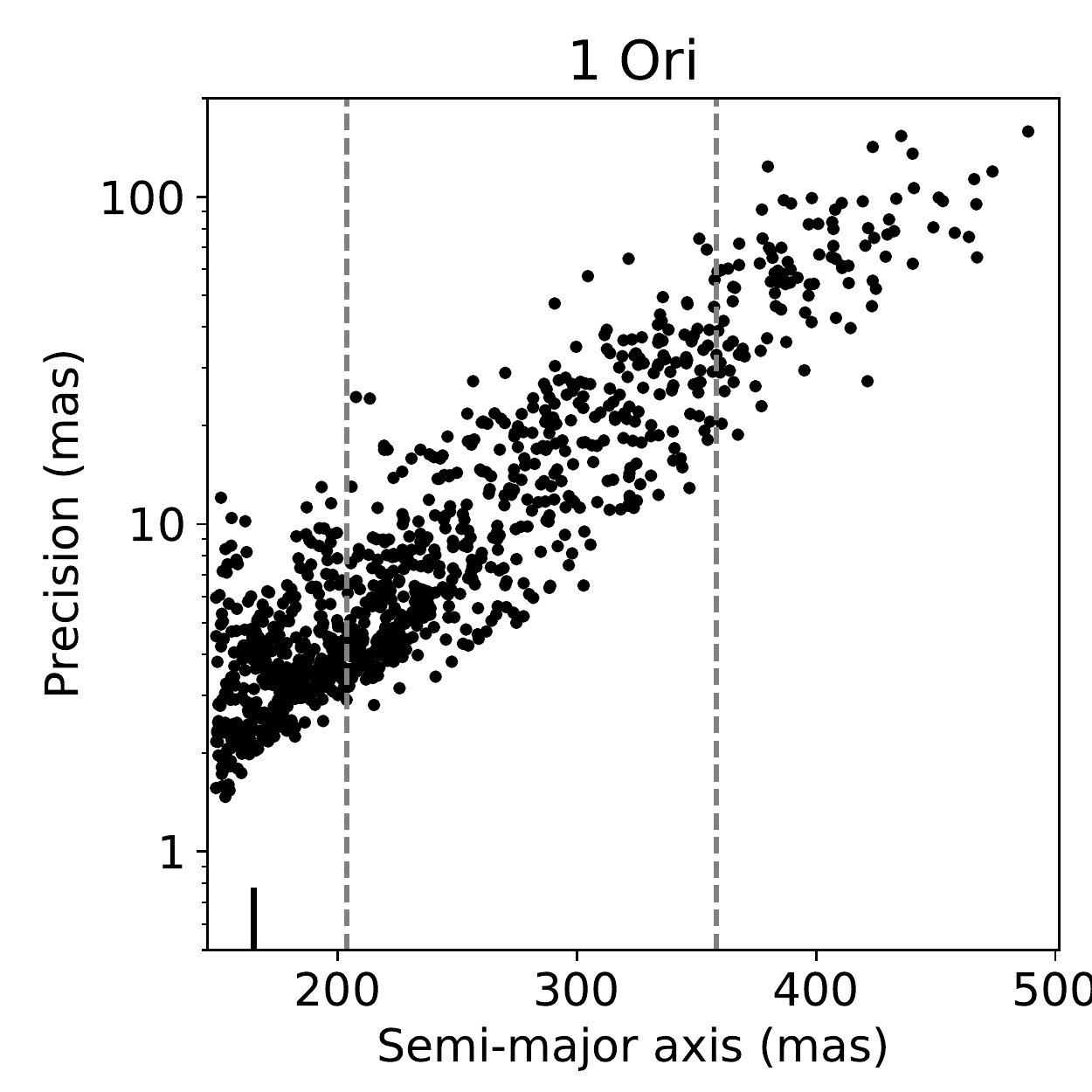}
     \includegraphics[width=2.in]{precision_summary.Fomalhaut.bw.pdf}
     \includegraphics[width=2.in]{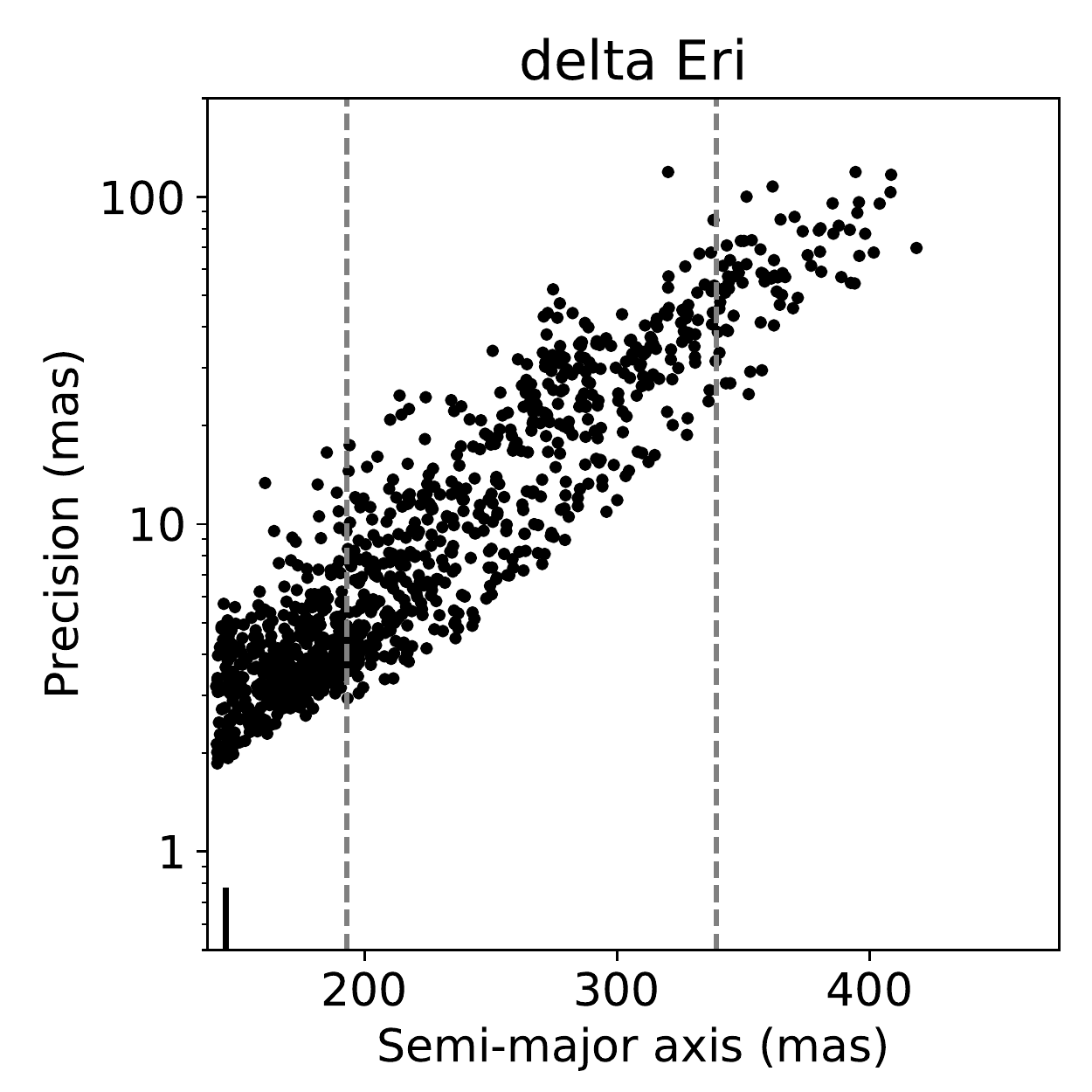}
     \includegraphics[width=2.in]{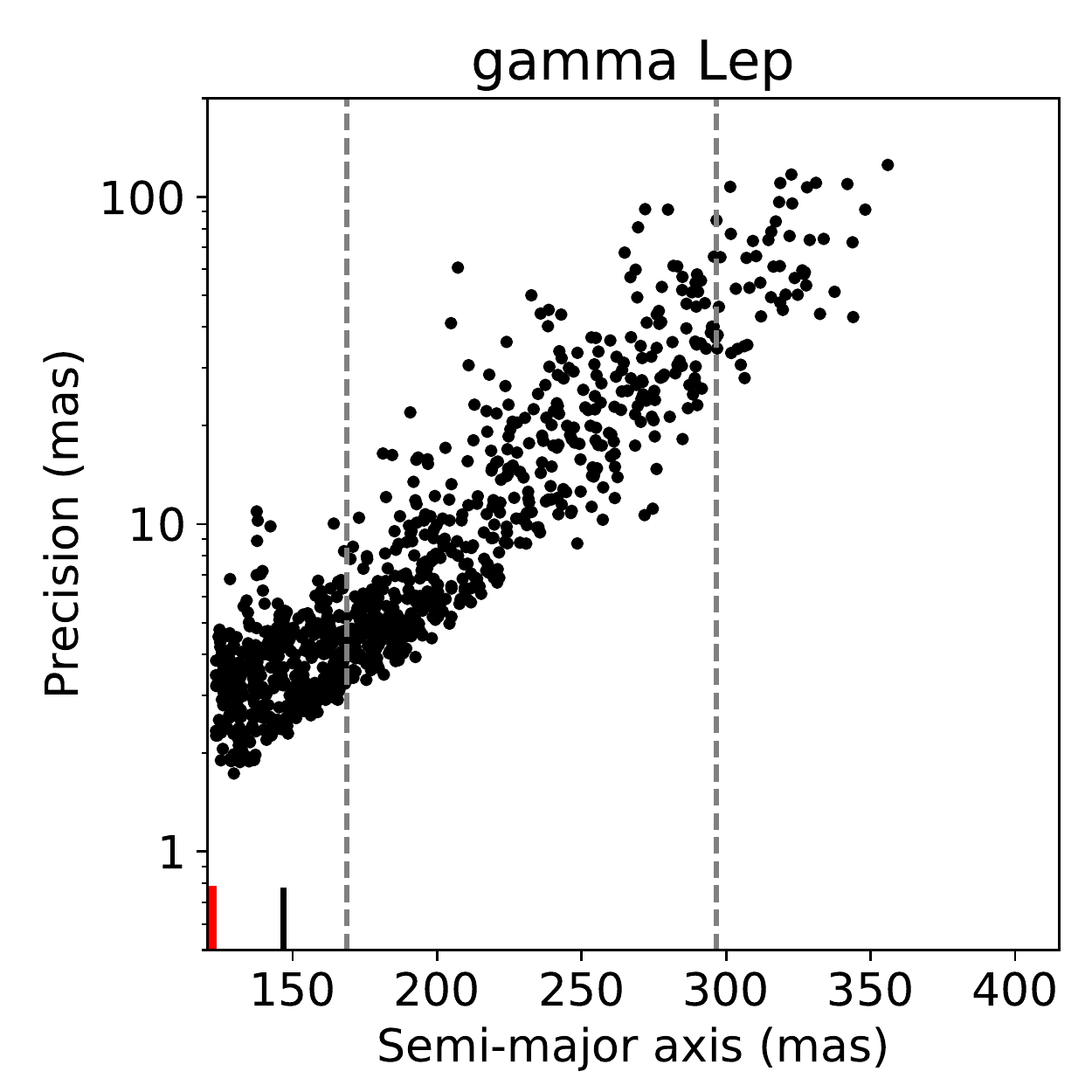}
     \includegraphics[width=2.in]{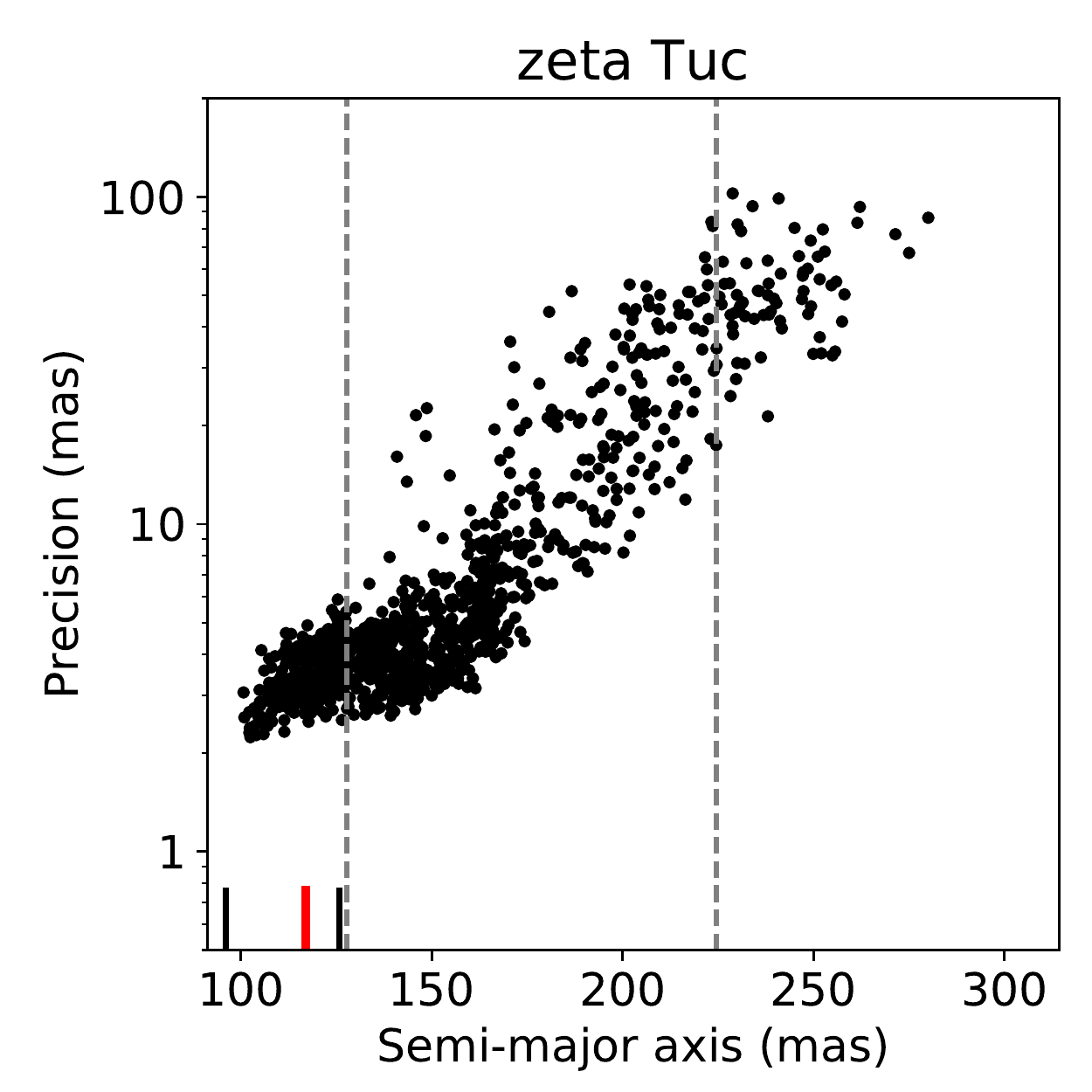}
  \end{center}
  \caption{
For each of our 16 target stars,
the orbital parameters are retrieved for 1000 random planet orbits,
each of which is directly imaged at least three times.
The precision for measuring each planet's semi-major axis is shown here.
  }\label{extraPrecisions}
\end{figure}

\begin{figure}[b]\begin{center}
     \includegraphics[width=2.in]{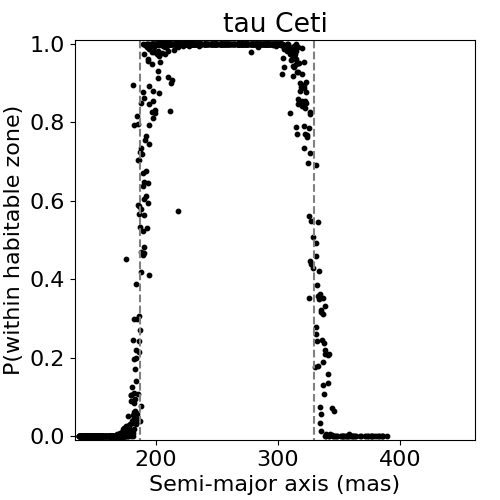}
     \includegraphics[width=2.in]{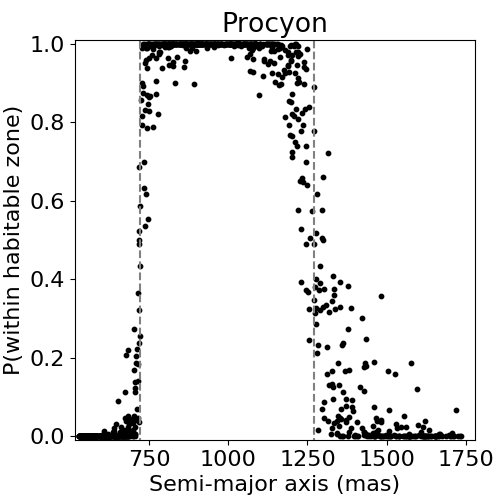}
     \includegraphics[width=2.in]{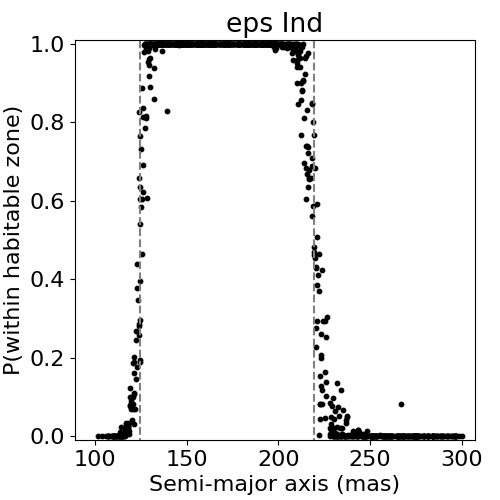}
     \includegraphics[width=2.in]{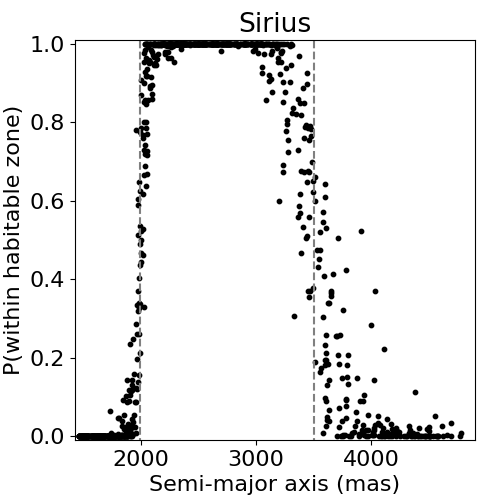}
     \includegraphics[width=2.in]{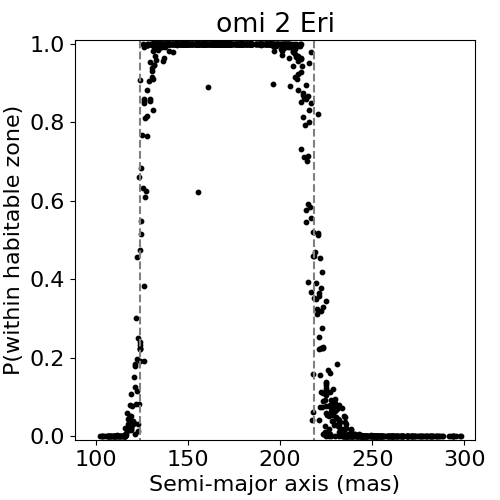}
     \includegraphics[width=2.in]{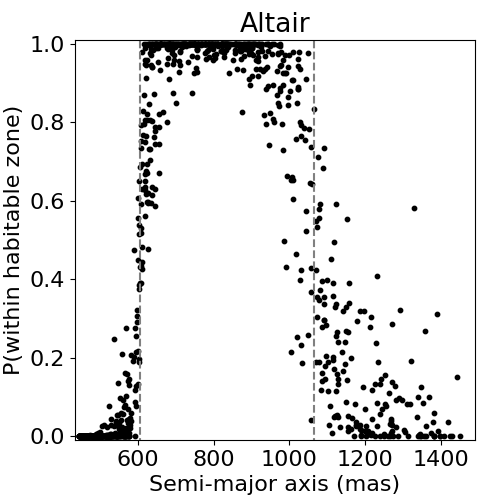}
     \includegraphics[width=2.in]{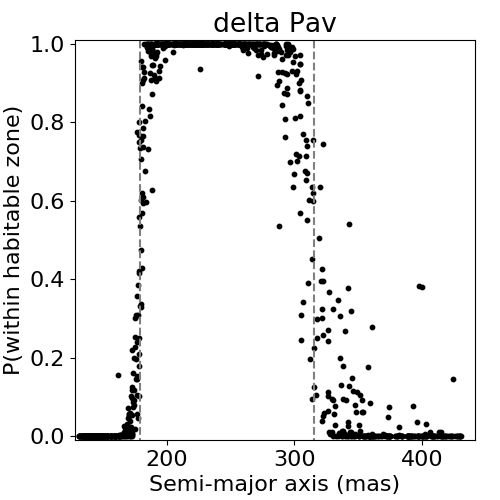}
     \includegraphics[width=2.in]{orbitfit_summary.82Eri.png}
     \includegraphics[width=2.in]{orbitfit_summary.sigmaDra.png}
  \end{center}\end{figure}\begin{figure}\begin{center}
     \includegraphics[width=2.in]{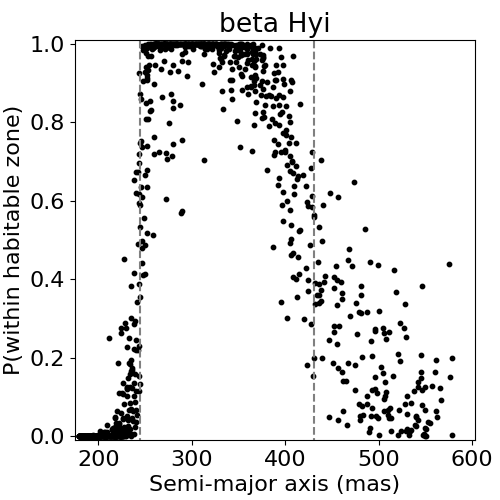}
     \includegraphics[width=2.in]{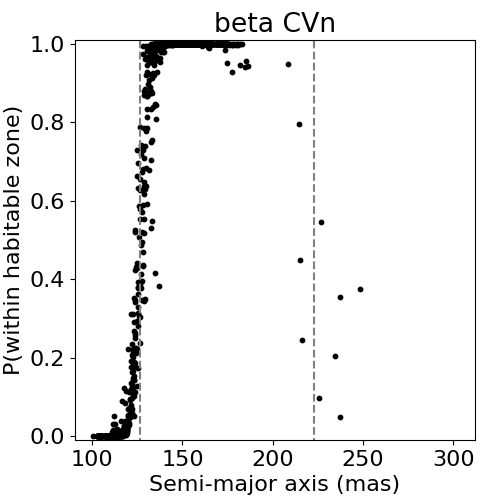}
     \includegraphics[width=2.in]{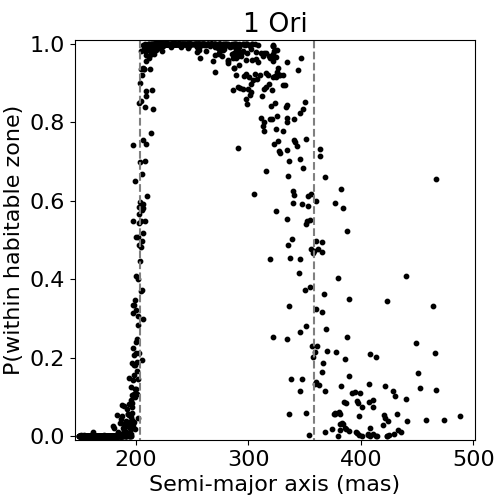}
     \includegraphics[width=2.in]{orbitfit_summary.Fomalhaut.png}
     \includegraphics[width=2.in]{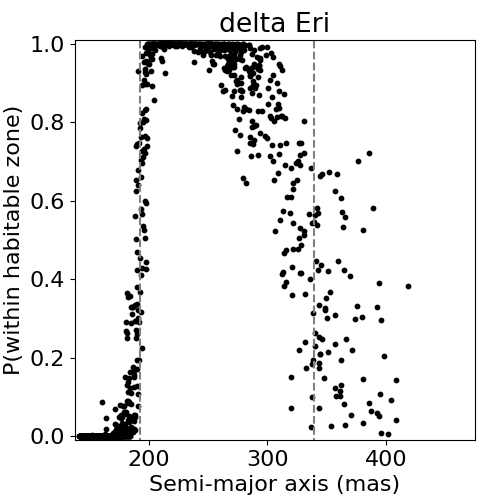}
     \includegraphics[width=2.in]{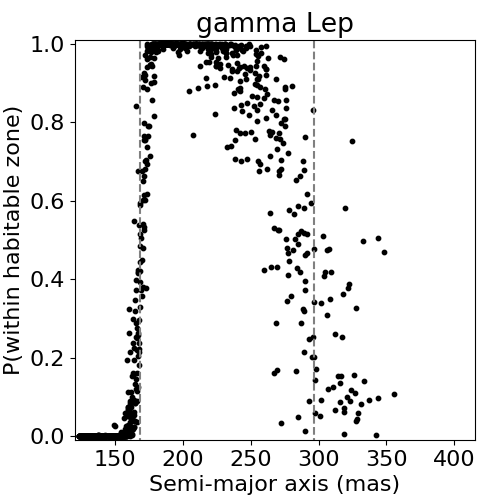}
     \includegraphics[width=2.in]{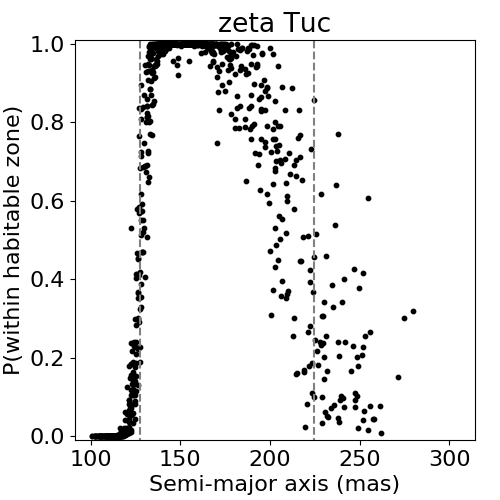}
  \end{center}
  \caption{
For each target star, orbital parameters and their uncertainties
are measured for 1000 random planetary orbits.
The probability of each planet residing in its star's habitable zone is
is shown as a function of its true semi-major axis.
  }\label{extraFig1}
\end{figure}


\end{document}